\documentclass[preprint,12pt]{aastex}

\newcommand\beq{\begin{equation}}
\newcommand\eeq{\end{equation}}

\begin{document}

\title{On the spin-up/spin-down transitions in accreting
X-ray binaries}

\author{Rosalba Perna\altaffilmark{1,}\altaffilmark{2}, 
Enrico Bozzo\altaffilmark{3}, Luigi Stella\altaffilmark{3}}

\affil{1. JILA and Department of Astrophysical and Planetary Sciences, 
University of Colorado, Boulder, CO, 80309}
\affil{3. INAF - Osservatorio Astronomico di Roma,
Via Frascati 33, I-00040 Rome, Italy}

\altaffiltext{2}{Also at Department of Astrophysical Sciences, Princeton
University, Princeton, NJ, 08544}

\begin{abstract}

Accreting X-Ray Binaries display a wide range of behaviours. Some of
them are observed to spin up steadily, others to alternate between
spin-up and spin-down states, sometimes superimposed on a longer trend
of either spin up or spin down.  Here we interpret this rich
phenomenology within a new model of the disk-magnetosphere
interaction. Our model, based on the simplest version of a purely
material torque, accounts for the fact that, when a neutron star is in
the propeller regime, a fraction of the ejected material does not
receive enough energy to completely unbind, and hence falls back into
the disk. We show that the presence of this feedback mass component causes the
occurrence of multiple states available to the system, for a given,
constant value of the mass accretion rate $\dot{M}_*$ from the
companion star. If the angle $\chi$ of the magnetic dipole axis with
respect to the perpendicular to the disk is larger than a critical
value $\chi_{\rm crit}$, the system eventually settles in a cycle of
spin-up/spin-down transitions for a constant value of $\dot{M}_*$ and
independent of the initial conditions.  No external perturbations are
required to induce the torque reversals. The transition from spin up
to spin down is often accompanied by a large drop in luminosity.  The
frequency range spanned in each cycle and the timescale
for torque reversals depend on
$\dot{M}_*$, the magnetic field of the star, the magnetic colatitude
$\chi$, and the degree of elasticity regulating the magnetosphere-disk
interaction. The critical angle $\chi_{\rm crit}$ ranges from $\sim
25^\circ-30^\circ$ for a completely elastic interaction to $\sim
40^\circ-45^\circ$ for a totally anelastic one.  For $\chi \la
\chi_{\rm crit}$, cycles are no longer possible and the long-term
evolution of the system is a pure spin up.  We specifically illustrate
our model in the cases of the X-ray binaries GX~1+4 and 4U~1626-67.

\end{abstract}


\section{Introduction}

Accreting X-ray binaries, with luminosities up to $\sim
10^{38}-10^{39}$ erg/s, constitute the brightest X-ray sources in the
sky. Since their discovery (Giacconi et al. 1971) more than three
decades ago, they have provided a unique laboratory to study, among
other things, the physical processes regulating accretion onto a
strongly magnetized neutron star (typically, $B\ga 10^{11}$ G).  In
neutron star (NS) binary systems containing a supergiant or a low-mass
star, mass transfer takes place through Roche Lobe overflow, and the
specific angular momentum of matter is sufficiently high to form an
accretion disk (the conditions for forming a disk are
however not necessarily met in Be-star systems, where the NS accretes
from capturing the star's wind; e.g. Rappaport 1982; Henrichs 1983).
In disk-fed systems that host a strongly
magnetic neutron star, the material from the disk is channelled toward
the magnetic poles, where it releases its gravitational energy giving
rise to the X-ray luminosity we observe. Pulsations at the
neutron star spin frequency are thus generated from a lighthouse-like
effect.  While the luminosity yields an estimate of the mass accretion
rate, pulse timing measurements allow one to measure the torque, and
hence probe the nature of the accretion process mediated by the
magnetosphere of the star. Early works (Pringle \& Rees 1972; Davidson
\& Ostriker 1973; Lamb, Pethick \& Pines 1973) showed that, when
accretion occurs through a prograde disk, the angular momentum
transferred by the accreting material to the star (material torque)
tends to spin the star up, until the centrifugal barrier inside the
corotation radius of the magnetosphere (Illianorov \& Sunyaev 1975)
becomes large enough to inhibit further accretion.  The star is then
expected to settle in a state with an equilibrium spin period which
depends mainly on the mass accretion rate provided by the companion
and the neutron star magnetic field (e.g. Frank, King \& Raine 2003).

Observations of disk-accreting X-ray pulsars during the 1970s and
1980s were rather sparse, and appeared to be roughly compatible with
the near-equilibrium picture (e.g. Nagase 1989), although there were
already hints at times of some unexpected behaviours. These included
torque reversals for some time while still accreting, or spin up rates
much smaller than expected for the observed luminosity.  In the 1990s,
continuous monitoring of several disk-fed X-ray pulsars with the Burst
and Transient Source Experiment (BATSE) on board the Compton Gamma-Ray
observatory, shed light on the long-term behaviour of several objects
(see Bildsten et al. 1997 for a comprehensive review).  Particularly
striking were the findings for GX~1+4 and 4U 1626-67: after about 15
years (for GX~1+4) and 20 years (for 4U 1626-67) of spin up, both
systems showed a torque reversal, which made them switch to a
spin-down phase.  Other systems, like Cen X-3, Vela X-1, Her X-1,
often showed an alternation of spin up and down sometime overimposed
on a longer term of either spin down or spin up.  In most cases, the
magnitude of the torques is comparable during the spin-up and the
spin-down regimes.  These unusual behaviours were a sign that the
simple scenario outlined above might be incomplete, and hence they
triggered a revival of research, mostly in the direction of finding
other sources of torque in addition to the one provided by the
accreting material alone.

Gosh \& Lamb (1979a,b, GL) and Wang (1987, 1995) suggested that, in
addition to the material torque, there is also an extra source of
torque provided by the magnetic field lines threading the disk. While
in the model of Pringle \& Rees (1972) the disk is truncated at the
point at which the magnetic pressure of the magnetosphere balances the
pressure of the accreting material, in the GL model there exists a broad
transition zone in which the magnetic field lines still thread the disk
even if the viscous stress in the disk material dominates over the magnetic
stress. This is made possible through the combination of a number of
effects, such as the Kelvin-Helmotz instability, turbulent
diffusion and reconnection.

In the models of Arons et al. (1984) and Lovelace et al. (1995) the
extra torque is provided by the expulsion of a magnetically-driven
wind.  Transitions between spin up and spin down states are possible,
but they must be induced by external perturbations, such as variations
in the viscosity parameter $\alpha$ of the disk or, most plausibly,
the accretion rate from the companion star.  These variations would
have to be finely tuned just so that the two torque states have
comparable magnitude but opposite sign. This seems unlikely in
general, but even more so in a system like 4U 1626-67, in which the
average mass accretion rate is likely determined by the loss of
orbital angular momentum via gravitational radiation (Chakrabarty et
al. 1997a).  Alternatively, in the case of GX 1+4, Makishima et
al. (1988) and Dotani et al. (1989) suggested that the spin down could
be due to accretion from a retrograde disk formed from the stellar
wind of the red giant companion. White (1988) however showed that this was
unlikely to be the case. A retrograde disk around the NS spin
axis could also be produced by magnetic torques generated in the
interaction between surface currents on the disk and the component
of the NS magnetic field parallel to the disk (Lai 1999).   

In this paper we discuss a new scenario for the spin up/spin down
transitions observed in binary systems accreting from a disk.  The
torque exchange between the magnetosphere and the disk material is
supposed to be dominated by the material component as in the early
models (Pringle \& Rees 1972). In this respect, our toy model is
very simple and idealized: possible torques non parallel to the
rotation axis are neglected, as well as magnetic torques (e.g. Gosh \&
Lamb 1979; Lai 2003).  What is new in our model is a computation of
the fate of the ejected material during the propeller phase of the
neutron star. Our calculation accounts for the following facts: {\em
i)} not all the ``propelled'' material receives sufficient energy to
unbind from the system; {\em ii)} if the magnetic moment of the
neutron star is inclined with respect to its rotation axis, there can
be, at the same time, regions of the magnetospheric boundary which are
allowed to accrete while others are propelling material away.  
This is a fundamental assumption of our model. While in this paper we
provide arguments in its support, a final validation will have to wait
for detailed numerical simulations. This work should therefore be
considered as an investigation (the first of its kind to the best of our knowledge) of the
characteristic timing behaviour of a pulsar whose magnetosphere can
simultaneously eject and accrete matter in different regions of its
boundary.  As it will be shown in the following, accounting for this
possibility leads to fundamentally different conclusions for the
long-term, equilibrium state of the system.  Rather than settling at
the equilibrium period at which the Keplerian frequency of the disk
matches the star rotation frequency at the point of interaction
(e.g. Frank, King \& Raine 1997), the system settles, for a wide range
of conditions, in alternating cycles of spin-up/spin-down for a
constant accretion rate from the companion star.  A qualitative
summary of our model is described below, and is formalized
mathematically in the following sections.

A magnetic neutron star surrounded by an accretion disk is able to
accrete only under the condition that the velocity of the
magnetosphere at the point of interaction (magnetospheric radius,
$R_M$) is smaller than the local Keplerian velocity of the disk
material.  If this condition is not satisfied, accretion is inhibited
(Illiaronov \& Sunyaev 1979), and angular momentum is transferred from
the star to the gas.  Whether this propelled gas can be completely
unbound from the system will depend on the location of the
magnetospheric radius within the gravitational field of the neutron
star. There exists a minimum distance, $R_{\rm inf}$, beyond which
ejection of matter to infinity is possible. If $R_M<R_{\rm inf}$, the
propelled material cannot be unbound, and therefore it will fall back
on the disk and accrete again. This matter is, in this sense,
``recycled''. An accreting system with recycled material can, under
certain conditions, have multiple states available.  This is due to
the fact that, for the system to be in a steady-state condition, the
total mass inflow rate at the magnetospheric boundary (which
determines the position of the magnetospheric boundary itself),
$\dot{M}_{\rm tot}= \dot{M}_{\rm acc} + \dot{M}_{\rm rec} +
\dot{M}_{\rm eje}$ must be such that $\dot{M}_{\rm acc} + \dot{M}_{\rm
eje}=\dot{M}_*$, where $\dot{M}_*$ is the mass inflow rate provided by
the companion star, and $\dot{M}_{\rm acc}\;,\dot{M}_{\rm rec}$ and
$\dot{M}_{\rm eje}$ are, respectively, the rate at which mass is
accreted, recycled and ejected.  Whenever the term $\dot{M}_{\rm rec}$
is non-negligible, there could be in principle different solutions to
the above condition corresponding to the same value $\dot{M}_*$ of the
mass inflow rate.  As the system spins up or down on a certain branch
of the solution, this solution can be lost, and the system is
consequently forced to jump to a different state, often characterized
by opposite torque.  This qualitative argument is formalized
mathematically in detail in \S2, while \S3 presents specific
applications to the cases of the accreting sources GX~1+4 and 
4U~1626-67.  Our results are summarized and discussed in \S4.

\begin{figure}[t]
\centering
\epsscale{.6}
\plotone{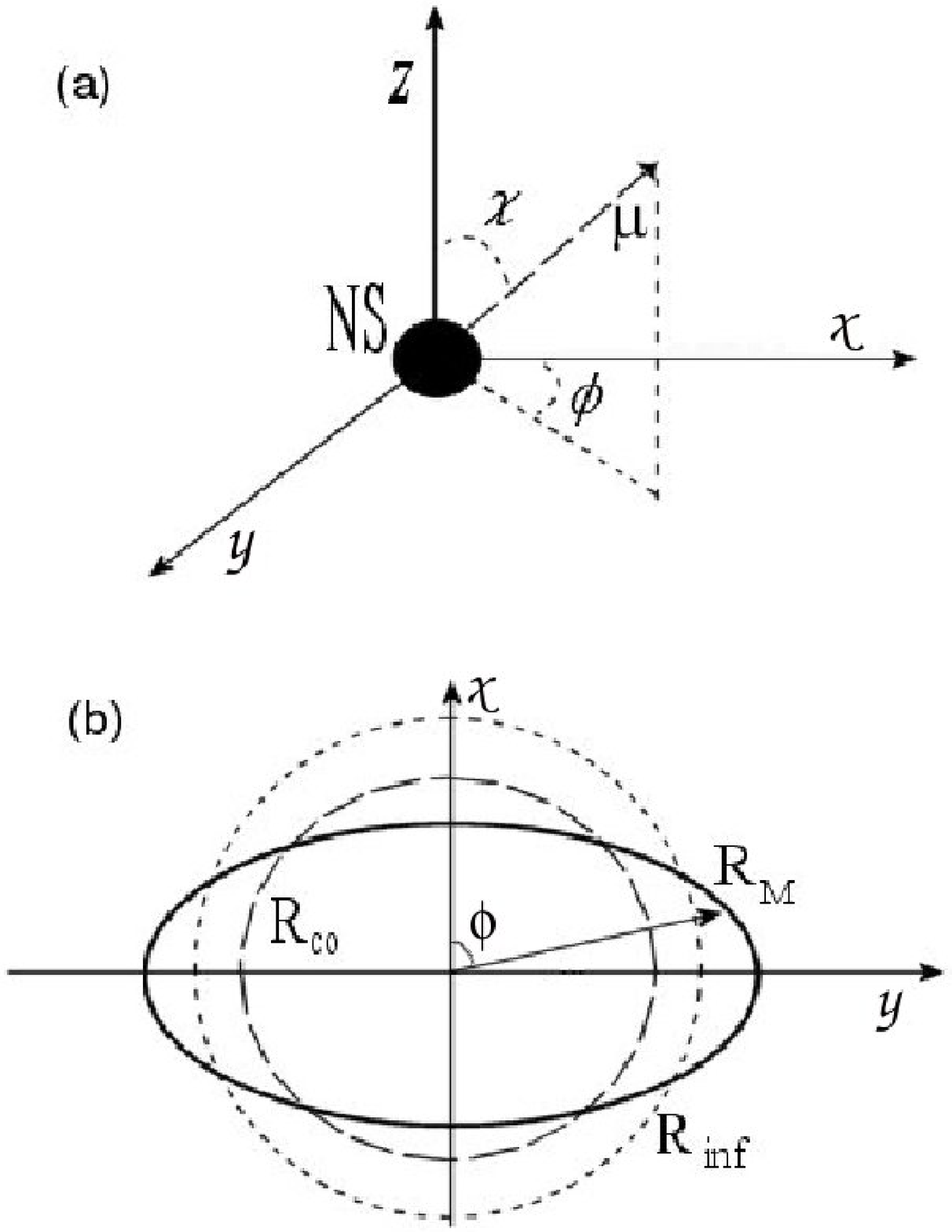}
\epsscale{1}
\caption{Schematic illustration of the
NS-disk system for an oblique rotator. Figure (a) shows the relative positions of
the magnetic dipole moment axis, the phase angle $\phi$, and the
inclination angle $\chi$. The NS is assumed to rotate around the
z-axis. Figure (b) is a two dimensional rapresentation of the position
of the magnetosphere (continuous line) with respect to the corotation
radius (long dashed line) and the infinity radius (short dashed line),
for arbitrarily fixed values of the NS parameters.}
\label{system}
\end{figure}

\section{Model description}

\subsection{Magnetosphere-disk interaction in an oblique rotator}

In this section we discuss the main concepts and assumptions upon
which our disk-magnetosphere model is based.  The basic geometry is
depicted in Figure \ref{system}. The axis of the magnetic moment
\textbf{$\mu$} of the neutron star (NS) is inclined with respect to
the rotation axis by the magnetic colatitude $\chi$. In cylindrical
coordinates $(r, \phi, z)$, where the z-axis coincides with the
rotation axis, the component of the magnetic field in the disk plane
is (Jetzer et al. 1998)
\begin{equation}
B^2=\frac{\mu^2}{r^6}\big[1+3(\sin{\chi}\sin{\phi})^2\big]\;,
\label{mag}
\end{equation} 
under the assumption that the disk is planar and its axis is parallel
to the spin axis of the NS.
When the rotation axis of the NS is inclined (i.e. $\chi\neq 0$), the
strength of the magnetic field in the plane of the disk depends on the
longitude $\phi$. As shown below, this angular dependence results
in an asymmetric magnetospheric boundary. 

The fate of the matter funnelled from the accretion disk to the rotating, magnetized
neutron star depends on a number of factors, the most 
important of which are the relative strength of the
magnetic pressure and the pressure of the accreting material, and the relative velocity
of the magnetosphere of the star at the inner radius of the disk with respect to
the Keplerian velocity at that same radius.
Following Lamb \& Pethick (1974), the former condition can be 
formalized by equating the magnetic energy
with the kinetic energy of the infalling matter:
\begin{equation}
\frac{1}{2}\rho v^2=\frac{B^2}{8\pi}\;.
\label{bal}
\end{equation}
In the free-fall approssimation the density is given by
$\rho=\rho_{ff}=\dot{M}/(4\pi v_{ff} r^2)$, where
$v_{ff}=(2GM/r)^{1/2}$ is the free-fall velocity. Using these
expressions, together with Eq.(\ref{mag}) and (\ref{bal}), the
magnetospheric radius for an oblique rotator can be obtained 
(Jetzer et al. 1998; see also Campana et al. 2001):
\begin{equation}
R_M(\phi)=3.2\times10^8\mu_{30}^{4/7}M_{1}^{-1/7}\dot{M}_{17}^{-2/7}
\left[1+3(\sin{\chi}\sin{\phi})^2\right]^{2/7}\;,
\label{mrad}
\end{equation}
where $\mu_{30}$ is the magnetic moment in units of
$10^{30}$~G~cm$^{3}$, $M_{1}$ is the NS mass in units of $1~M_{\odot}$
and $\dot{M}_{17}$ is the accretion rate in units of
$10^{17}$~g~s$^{-1}$. The minimum radius $R_{\rm M}(0)$ also corresponds
to $\chi=0$, the approximation usually adopted in models of the 
disk-magnetosphere interaction. The maximum radius $R_{\rm M}(\pi/2)$
is only a factor of $(1+3\sin^2\chi)^{2/7}\le 1.49$ larger. 
Note that the elongated shape of the magnetospheric
boundary plays a fundamental role in our model.

An important assumption of our model is that, during the rotation of
the magnetosphere (whose shape depends upon the instantaneous position
of the magnetospheric radius as a function of $\phi$), matter in the
Keplerian disk is able to fill the region that separates the disk and
the magnetospheric flow on a timescale shorter than the spin period of
the star.  This ensures that the inner boundary of the disk remains in
constant contact with the magnetosphere.  We show in the appendix that
the Kelvin-Helmholtz instability operates on a sufficiently short
timescale and wide range of radii that this assumption can be
justified.

Accretion to the star is possible only under the condition that, at
the magnetospheric radius, the Keplerian velocity of the accreting
gas, $\Omega_{\rm K}(R_M)$, is larger than the velocity $\Omega_0$ of
the rotating magnetosphere of the star (equal to the velocity of the
star), otherwise centrifugal forces will inhibit accretion (Illianorov
\& Sunyaev 1979).  The above condition is equivalent to saying that the
magnetospheric radius must be smaller than the corotation radius,
$R_{\rm co}=(GM/\Omega_0^2)^{1/3}$, which is the radius at which the
Keplerian frequency of the orbiting matter is equal to the NS spin
frequency $\Omega_0$. In an oblique rotator, the onset of the
propeller stage will occur when $R_M(\phi)=R_{\rm co}$ at least in one
point of the magnetospheric boundary. Note that, while a parallel rotator can be
{\em either} in the propeller {\em or} in the accreting regime, an
oblique rotator can be in both states {\em simultaneously} for
different longitudes of the magnetospheric boundary. Indeed, this
special feature of the oblique rotator was used by Campana et
al. (2001) in building up a model that explained the dramatic
luminosity variations seen in the {\em BeppoSAX} observation of the
transient X-ray pulsar 4U 0115+63\footnote{This simultaneous presence of different regimes, which is crucial
to our model, has not yet been seen in numerical simulations. However, to
the best of our knowledge, current numerical simulations of the propeller regime
(e.g. Romanova et al. 2004) are axisymmetric; because of this geometry, they
cannot verify the simultaneous presence of different regimes of the kind discussed here.}.

The interaction between the magnetosphere of the NS and the matter in the
disk is likely to be at least partially anelastic because of 
dissipative effects in the mixing process between the
magnetospheric plasma and the disk matter during the propeller phase.  For
clarity of presentation, here we consider first the two limiting cases
of a completely anelastic and a completely elastic interaction, and
then we will generalize our results to the partially anelastic case.

In the anelastic case, the magnetic field of the NS is able to force matter
to corotate at the same velocity of the star, and it is endowed at the
magnetospheric boundary with specific kinetic energy
$\epsilon=1/2\Omega^2_0 R_M^2$ and angular momentum $l=\Omega_0 R_M^2$.
In order for matter
to be ejected from the system via the propeller mechanism, the
magnetic field must provide it with enough energy to reach a velocity in excess
of the local escape velocity at $R_M$. Because in the anelastic case
the ejection velocity is $v_{ej}=\Omega_0 R_M$, the requirement above
converts to an ''ejection radius''
\begin{equation}
R_{\rm inf, ane}=(2GM/\Omega_0^2)^{1/3}\simeq1.26 R_{\rm co}\;. 
\label{Rinf}
\end{equation}   
Only matter which is located beyond this radius during the interaction
with the magnetosphere of the NS can be unbound from the system through
the propeller mechanism. Therefore there exists a region 
($R_{\rm co}<R_{M}<R_{\rm inf}$) in which the propeller is active but matter cannot
be unbound from the sytem by merging with the disk matter (Spruit \& Taam, 1993). We
assume that matter in this zone is swung out and circularizes at the
radius where its angular momentum equals the Keplerian value,
i.e. when $l=\Omega_0 R_M^2=l_{\rm K}=\Omega_{\rm K}(R_{\rm K}) R_{\rm K}^2$ (here $R_{\rm K}$ is
the circularization radius). This condition defines the
Keplerian circularization radius:
\begin{equation}
R_{\rm K,{ane}}=\frac{\Omega_0^2 R_M^4}{GM}\;.
\label{circ}
\end{equation} 
Matter that is not ejected from the system will fall back into the
disk and restart its motion toward the NS from the radius defined in
Equation (\ref{circ}).

In the case of the elastic propeller, we assume that
material in the disk at $R_M$ moves toward the magnetosphere with a
tangential relative velocity of $-v_{\rm rel}=R_M(\Omega_0-\Omega_{\rm K})$,
where $\Omega_{\rm K}$ is the Keplerian angular velocity at $R_M$. In a
completely elastic interaction this matter bounces off at the
magnetospheric boundary with an
opposite velocity of $+v_{\rm rel}$ that in the non-rotating frame sums
with $v_{\rm rot}=\Omega_0 R_M$. Thus the ejection velocity is
$v_{ej}=R_M[2\Omega_0-\Omega_{\rm K}(R_M)]$. In this case the requirement
that this velocity be larger than $ v_{esc}(R_M)$ can be
written as:
\begin{displaymath}
\frac{R_M^2}{2}\Big[4\frac{GM}{R_{\rm co}^3}+\frac{GM}{R_M^3}
-4\frac{GM}{R_{\rm co}^{3/2}R_M^{3/2}}\Big]\geq\frac{GM}{R_M}
\end{displaymath} 
where we have used the definition of the corotation radius. This
equation can be solved as a function of the magnetospheric radius to
define the limit beyond which ejection of matter to infinity 
is possible in the purely elastic case:
\begin{equation}
R_{\rm inf,{el}}=\Big[\frac{1+\sqrt{2}}{2}\Big]^{2/3}R_{\rm co}\simeq1.13 R_{\rm co}\;.
\label{Rinfel}
\end{equation}
The matter leaving the magnetospheric boundary is endowed with
specific angular momentum $l_{el}=R_M^2(2\Omega_0-\Omega_{\rm K})$; equating
this to $l_{\rm K}$ gives a new circularization radius for matter that is
not ejected to infinity. Using the same notation as above we find:
\begin{equation}
R_{\rm k,{el}}=\frac{R_M^4(2\Omega_0-\Omega_{\rm K})^2}{GM}\;.
\label{Rkel}
\end{equation}
Let us now consider the most general case of a partially elastic
interaction. Following the formalism developed by Eksi et al. (2005),
we define the ``elasticity parameter'' $\beta$, which is a measure of
how efficiently the kinetic energy of the neutron star is converted
into kinetic energy of ejected matter through the magnetosphere-disk interaction.
Taking into account
the definitions given above, we now consider the generalized
rotational velocity of matter at the magnetospheric boundary:
\begin{equation}
v_{\rm gen}=\Omega_{\rm K}(R_M)R_M(1-\gamma)
\label{vgen}
\end{equation}
where $\gamma=(1+\beta)(1-\Omega_0/\Omega_{\rm K})$. 
The elastic case is obtained in the limit $\beta=1$, and the totally 
anelastic one when
$\beta=0$. Using Equation (\ref{vgen}) we can then generalize also the
expression for the infinity radius
\begin{equation}
R_{\rm inf}=\Big(\frac{\beta+\sqrt{2}}{1+\beta}\Big)^{2/3} R_{\rm co}
\label{rifg}
\end{equation} 
and for the circularization radius
\begin{equation}
R_{\rm K}=R_M (1-\gamma)^2\;.
\end{equation}
In our model we will consider the general case of a partially elastic
interaction, and use $\beta$ as one of the model parameters.

Figure \ref{system} (b) shows the various characteristic radii defined
above on the disk plane $z=0$. 
Depending on the phase ($\phi$) and 
the inclination angle ($\chi$), it is possible to have regions of the
magnetospheric boundary in which accretion is possible ($R_M(\phi,
\chi)<R_{\rm co}$) together with other portions in which the propeller is
already active, resulting in ejection of matter to larger radii
($R_{\rm co}<R_M(\phi, \chi)<R_{\rm inf}$), or to infinity ($R_M>R_{\rm inf}$). In
those cases in which the inclination angle is sufficiently large, it
is possible to have  all the three regimes described
above simultaneously.

It should be noted that, in our model, we consider ejection of matter
from regions of the disk that are away from the corotation radius,
where the Keplerian velocity of matter becomes rapidly supersonic
(e.g. Frank, King, Raine 2003). This could in principle lead to the
formation of supersonic shocks which can heat the plasma and
eventually stop the ejection mechanism. However in this situation, due
to the high relative rotation rate between the plasma inside the
magnetosphere and that inside the disk, the Kelvin-Helmholtz
instability can be very efficient. As previously discussed, this
instability can lead to a large mixing of the two fluids, providing a
mechanism to mantain the interaction between the magnetic field of the
NS and the matter in the disk. Under these circumstances, it has been
shown that outflowing bubbles of matter are likely to be accelerated
magnetically by the NS towards the outer region of the disk (Wang \&
Robertson 1985), in turn supporting the idea that ejection
far away from the corotation radius can be sustained.

\begin{figure}[t]
\centering
\plotone{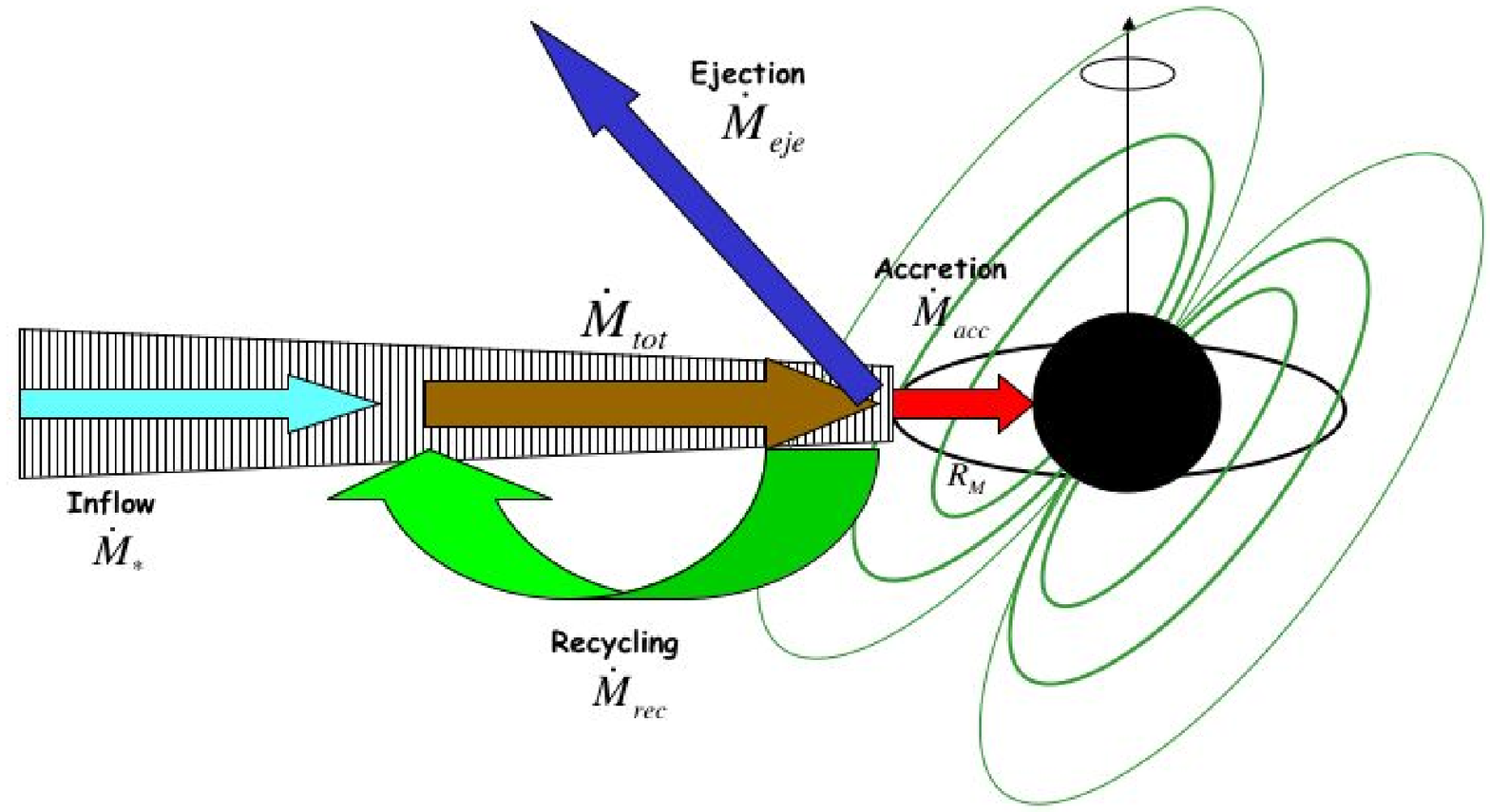}
\caption{The fate of the matter provided by the companion at a rate $\dot{M}_*$
depends on the relative position of the magnetospheric radius with
respect to the corotation radius and the ejection radius. Matter can
be accreted, ejected or recycled into the disk.}
\label{mdotgeom}
\end{figure}

\subsection{Conditions for the existence of a limit cycle}

\begin{figure}[t]
\centering
\plotone{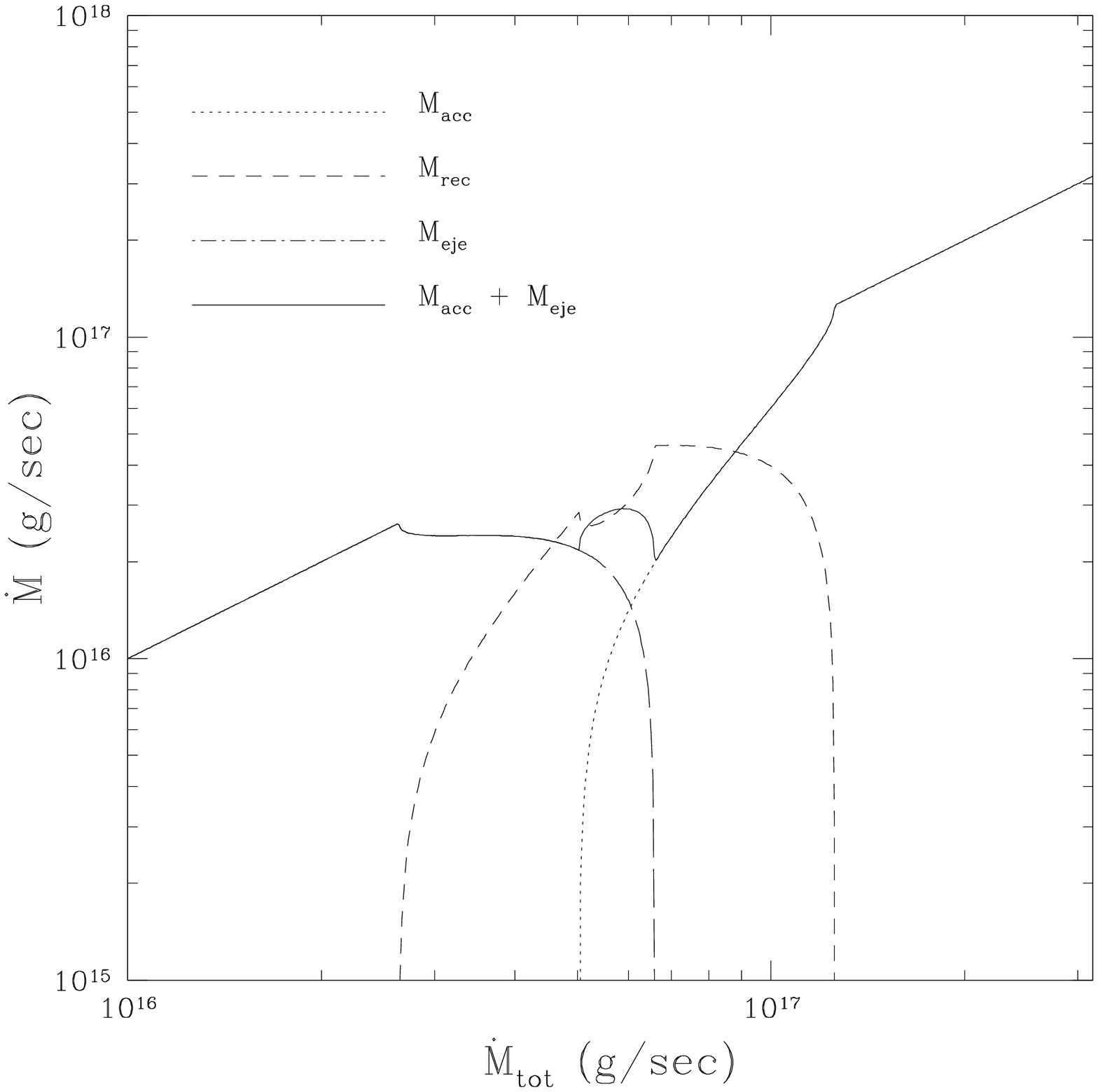}
\caption{The various contributions to the total accretion
rate  $\dot{M}_{\rm tot} = \dot{M}_{\rm acc} + \dot{M}_{\rm rec} + \dot{M}_{\rm eje}$
at the magnetospheric-disk boundary. The system parameters are $B=6\times 10^{13}$ G,
$\nu=9$ mHz, $\chi=45^\circ$, $\beta=0.3$.}
\label{mdot}
\end{figure}

Let $\dot{M}_*$ be the rate of inflowing matter, regulated through the
Roche Lobe overflow or capture of part of the wind of the companion
star.  We assume that this matter possesses in all cases enough
angular momentum that a prograde accretion disk forms. We further
assume that the mass inflow at the inner disk boundary is azimuthally
symmetric (i.e. independent of $\phi$).  As illustrated in Figure 1
and discussed in \S 2.1, for a general, oblique, orientation of the
magnetic field of the NS with respect to the normal to the disk and
the spin axis of the NS (which we assume are parallel), there will be
regions where $R_M(\phi)<R_{\rm co}$, and therefore some matter is
able to accrete, regions for which $R_M(\phi)>R_{\rm inf}$ that result
in matter being ejected, and intermediate zones with $R_{\rm
co}<R_{M}(\phi)<R_{\rm inf}$ from which matter gets ``recycled''. The
fraction of material in each of these regions is expected to be
proportional to the angle $\phi$ subtended by the relevant region in
the magnetosphere as shown in Figure 1. As in \S1, let us define
$\dot{M}_{\rm acc}$, $\dot{M}_{\rm eje}$ and $\dot{M}_{\rm rec}$ to be
respectively the rates of accreting, ejected and recycled material at
any given time. These various components are illustrated in Figure
\ref{mdotgeom}. If $d\dot{M}_{\rm tot}/d\phi$ is the total rate of
matter exchanged at the magnetosphere-disk boundary per unit angle,
these components are given by $\dot{M}_{\rm comp}=1/2\pi\int_{\phi_1}^{\phi_2} d\phi 
\left(d\,\dot{M}_{\rm tot}/d\phi\right)$, where the integration interval $[\phi_1,\phi_2]$
of $\phi$ is such that $R_M(\phi)<R_{\rm co}$ when comp=''acc'' , 
$R_{\rm co}<R_{M}(\phi)<R_{\rm inf}$ when comp=''rec'' and 
$R_M(\phi)>R_{\rm inf}$ when comp=''eje''. 
Figure \ref{mdot} shows an example of these components as a function of
the total mass inflow across the entire magnetospheric boundary,
$\dot{M}_{\rm tot}$. At low values of $\dot{M}_{\rm tot}$, $R_{\rm inf}
>R_M$ for all values of $\phi$, and therefore all matter is ejected
 (i.e. $\dot{M}_{\rm eje}=\dot{M}_{\rm tot}$). On the other hand, at
high values of $\dot{M}_{\rm tot}$, $R_{\rm co}
>R_M$ for any $\phi$, and therefore all matter is accreted 
($\dot{M}_{\rm acc}=\dot{M}_{\rm tot}$).
For values of $\dot{M}_{\rm tot}$ such that $R_M(\phi)$ crosses
$R_{\rm co}$ at some values of $\phi$, $\dot{M}_{\rm rec}\neq 0$. 

While the total mass inflow rate available to the system is determined
by the mass transfer rate from the companion, $\dot{M}_*$, the value
of the magnetospheric radius $R_M$, on the other hand, is 
determined by the total pressure of the accreting matter,
i.e. $\dot{M}_{\rm tot}=\dot{M}_{\rm acc}+\dot{M}_{\rm
eje}+\dot{M}_{\rm rec}$. Since, in general, $\dot{M}_{\rm tot}\ge\dot{M}_*$,
the magnetospheric radius can be smaller than it would be
if the ``recycled'' mass component were not accounted for (as
commonly assumed in the literature). Therefore, including 
$\dot{M}_{\rm rec}$ in the computation of $R_{M}$, allows
accretion at the same rate to occur for smaller values of $\dot{M}_*$ than
it would otherwise. 

\begin{figure}[t]
\centering
\plotone{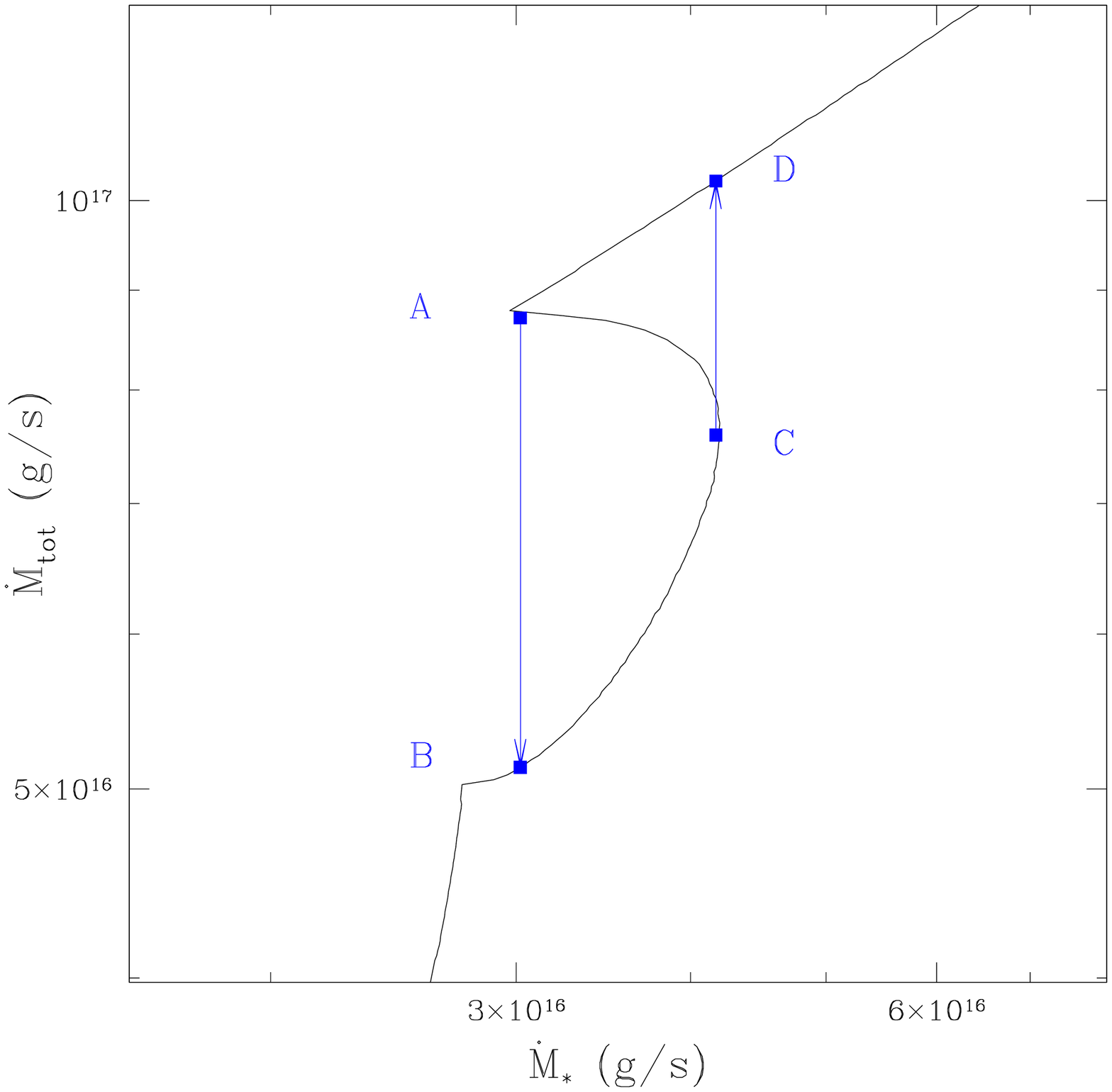}
\caption{Schematic representation of the hysteresis-type limit cycle.
The arrows indicate the points where the
system ''jumps'' between different states as a result of variations
in the external mass supply rate $\dot{M}_*$. 
The system parameters in this example are $\nu=9$ mHz,
$B=6\times 10^{13}$ G, $\chi=80^\circ$, $\beta=0$.}
\label{timeindependent}
\end{figure}

In order to demonstrate the existence of a limit cycle,
testified by a hysteresis-like
loop in the $\dot{M}_* - \dot{M}_{\rm tot}$ plane, we start by noting
that the rate at which matter is ``recycled'', $\dot{M}_{\rm rec}$, 
does not contribute to the mass budget; therefore
a steady-state solution is possible only if
\beq
\dot{M}_{*}=\dot{M}_{\rm acc}+\dot{M}_{\rm eje}\;.
\label{eq:sol}
\eeq 
Let us therefore examine the behaviour of
the curve $\dot{M}_{\rm tot}$ as a function of the accretion rate
$\dot{M}_{\rm acc}+\dot{M}_{\rm eje}=\dot{M}_*$.  An example of such a
curve for a rotator inclined by an angle of $\chi=50^\circ$ is shown
in Figure \ref{timeindependent}.  All the characteristic parameters of
the NS ($B$, $\Omega_0$, $R_{NS}$, $M_{NS}$) and the 
angle $\chi$ are kept fixed while the mass supply from the companion
is varied. For a given value of the external rate of mass supply
$\dot{M}_*$, the corresponding points on the curve yield the value (or
values) of $\dot{M}_{\rm tot}$ for which there exists a solution.
Again, we stress that the ``state'' of the system, and the characteristics
of the solution, are determined by $\dot{M}_{\rm tot}$ since it is 
this quantity (and not $\dot{M}_*$) which determines the position of $R_M$. 
There can be multiple solutions for a given $\dot{M}_*$, and
the one that is realized at a certain time depends on the previous
history of the system. This situation is reminiscent of a system with
hysteresis, and in fact, as Figure \ref{timeindependent} shows, the
shape of the curve $\dot{M}_{\rm tot}(\dot{M}_*)$ resembles a
hysteresis curve, where the role of the external magnetic field is
played by the rate of mass supply by the companion, $\dot{M}_*$ (the
independent variable in the present context).  If at a certain point
the system is in, say, the state indicated by the point ``C'' in the
figure, and $\dot{M}_*$ increases, the solution (i.e. only available
state for the system) will be forced to jump to the state indicated by
point ``D''. As $\dot{M}_*$ decreases, the solution will move from
``D'' to ``A''  but from that point on, any further
decrease in $\dot{M}_*$ will cause the solution to jump to point ``B''.
Therefore, like in the traditional
hysteresis cycle, continuous variations in $\dot{M}_*$ result in
discontinuous states for the system.

In the following section, after discussing the computation of the
torque, it will be shown that the points where the solution jumps from
one place to another in the $\dot{M}_{\rm tot}(\dot{M}_*)$ curve often
straddle the point of torque reversal. Therefore, transitions
between different states are often characterized by a torque reversal. 

The case we have illustrated in Figure~\ref{timeindependent} is only
an example of a cyclic behaviour. The shape of the curve 
$\dot{M}_{\rm tot}(\dot{M}_*)$ changes with the parameters
$\chi$ and $\beta$ (while $\nu$ and $B$ only cause a translation in 
the $\dot{M}_{\rm tot}-\dot{M}_*$ plane). This can result
in several types of cycles with a different number of jumps.
More examples are shown in Figure \ref{cycles}.

\subsection{Torque and luminosity in the different states of an oblique rotator}

We calculate here the net specific angular momentum trasferred between
the disk and the NS.  In the region of the magnetospheric boundary
where accretion is allowed, the net specific angular moment
transferred from the disk to the NS is given by
\begin{equation}
l_{\rm acc}=\frac{1}{2\pi}\int_{R_M<R_{\rm co}}(GMR_{M})^{1/2}d\phi\;.
\label{lacc}
\end{equation} 
In the ejection region, the NS accelerates the material to the
ejection velocity, which, as discussed in \S 2.1, is different in the two limiting cases of
a completely elastic or anelastic propeller. For the general
case of a partially elastic interaction, using Equation (\ref{vgen}),
the angular momentum given by the NS to the ejected matter is:
\begin{eqnarray}
\lefteqn{l_{eje}=\frac{1}{2\pi}\int_{R_M>R_{\rm co}} (v_{gen}R_M-\Omega_{\rm K}R_M^2) d\phi={}}
\nonumber \\
& & {}\frac{1}{2\pi}\int_{R_M>R_{\rm co}} \Omega_{\rm K} R_M^2(1+\beta)(\Omega_0/\Omega_{\rm K}-1) d\phi\;.
\label{lej}
\end{eqnarray}
By relating this transfer of
angular momentum between the NS and the disk to the variation of the
NS angular momentum, we have
\begin{equation}
\frac{d\Omega_0}{dt}=\frac{\dot{M}_{\rm tot}l_{\rm tot}}{I}
\label{angm}
\end{equation}
where $l_{\rm tot}$ is the sum of the angular momentum computed from
Equations (\ref{lacc}) and (\ref{lej}), and we have assumed that the
variation of the NS moment of inertia ($I$) is negligible. Using
equations (\ref{lacc}) and (\ref{lej}) in (\ref{angm}), we obtain:
\begin{equation}
\frac{d\Omega_0}{dt}=\frac{\dot{M}_{\rm tot}}{2\pi}\int_{0}^{2\pi}
(GMR_M)^{1/2}\big\{1-\theta(R_M-R_{\rm co})\big[(1+\beta)(\Omega_0/\Omega_{\rm K}-1)\big]\big\} d\phi
\end{equation}
where $\theta(R_M-R_{\rm co})$ is $1$ for $R_M>R_{\rm co}$ and $0$ for
$R_M<R_{\rm co}$.

\begin{figure}[t]
\centering
\plottwo{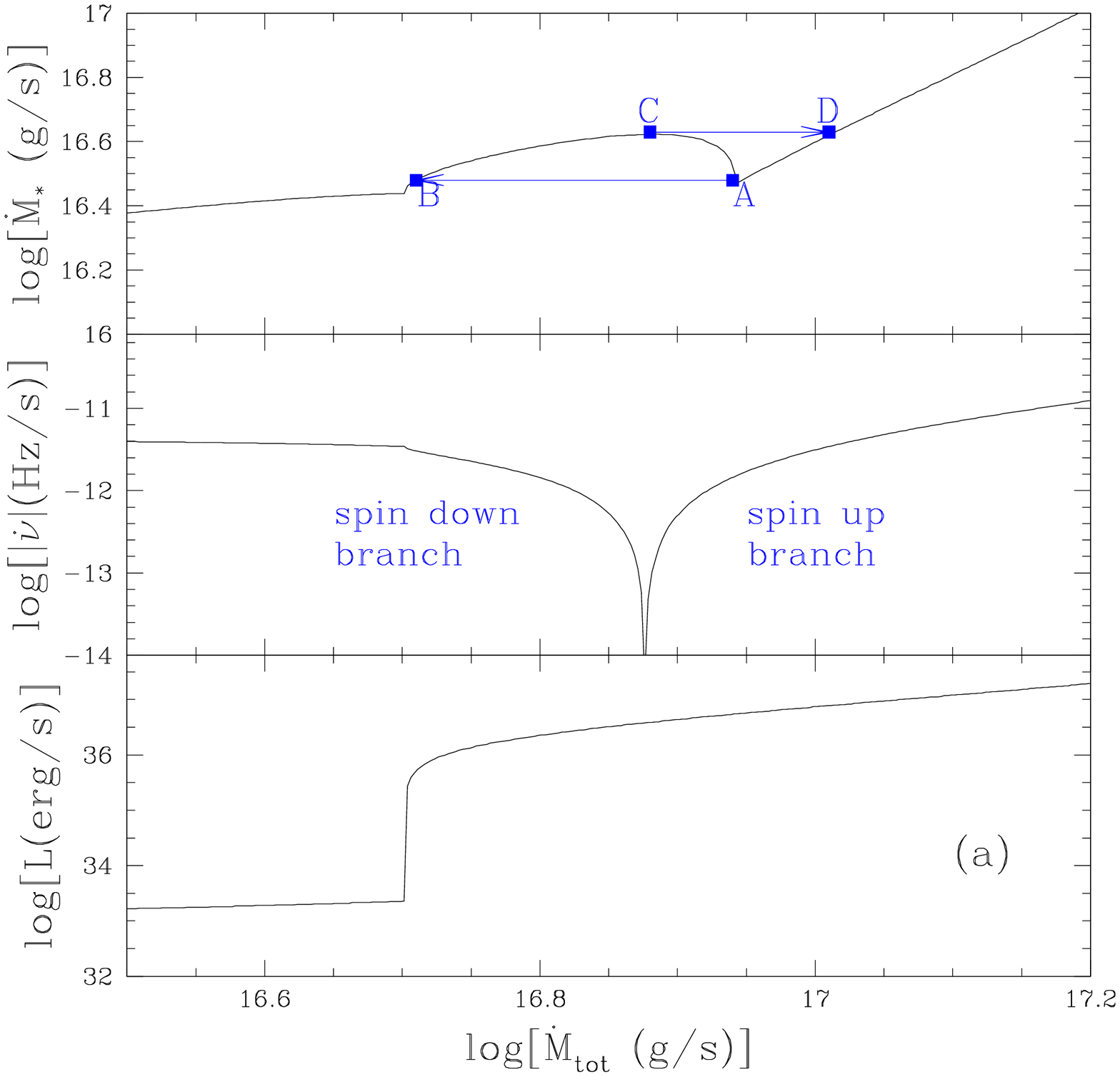}{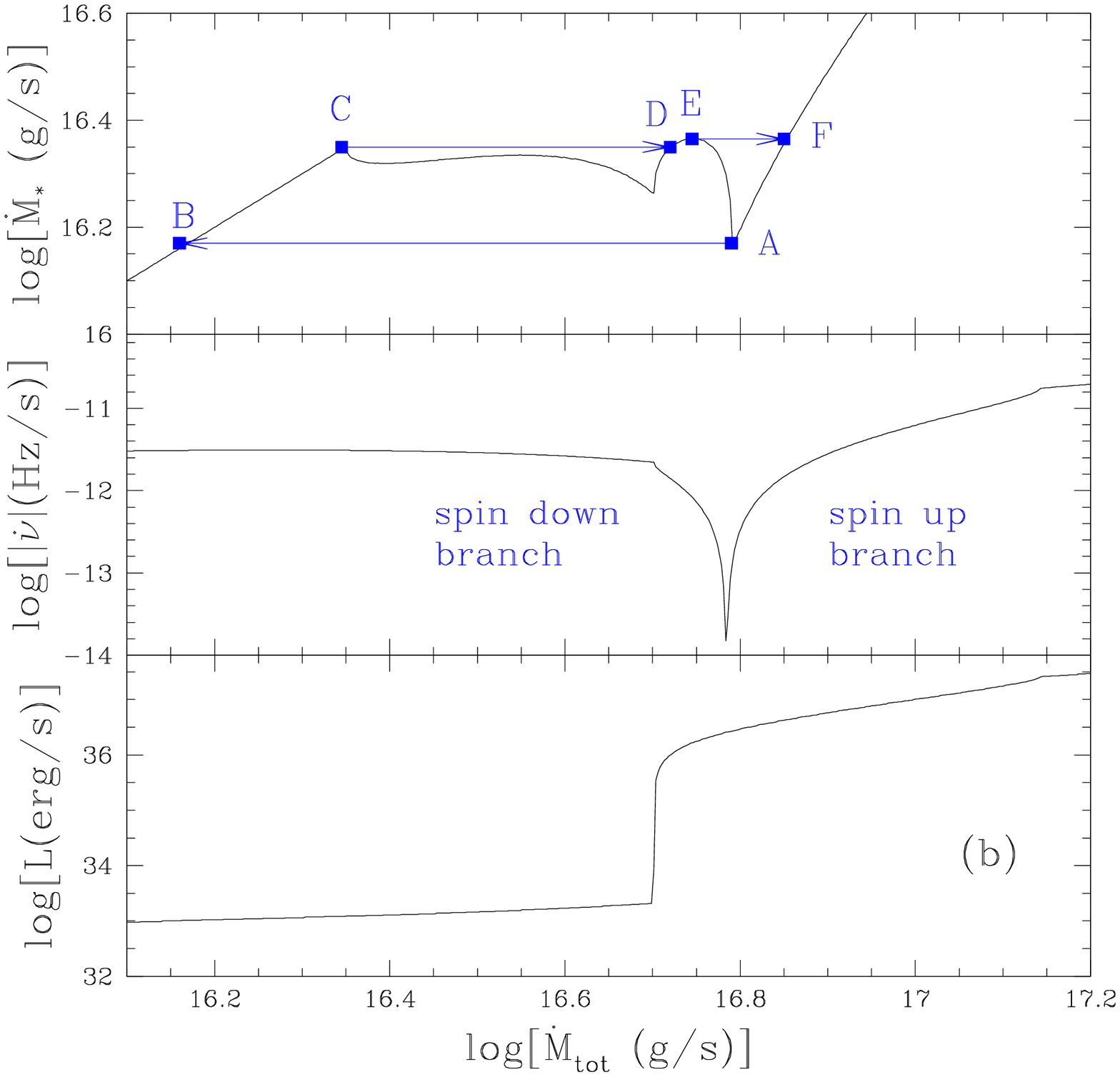}
\epsscale{.5}
\plotone{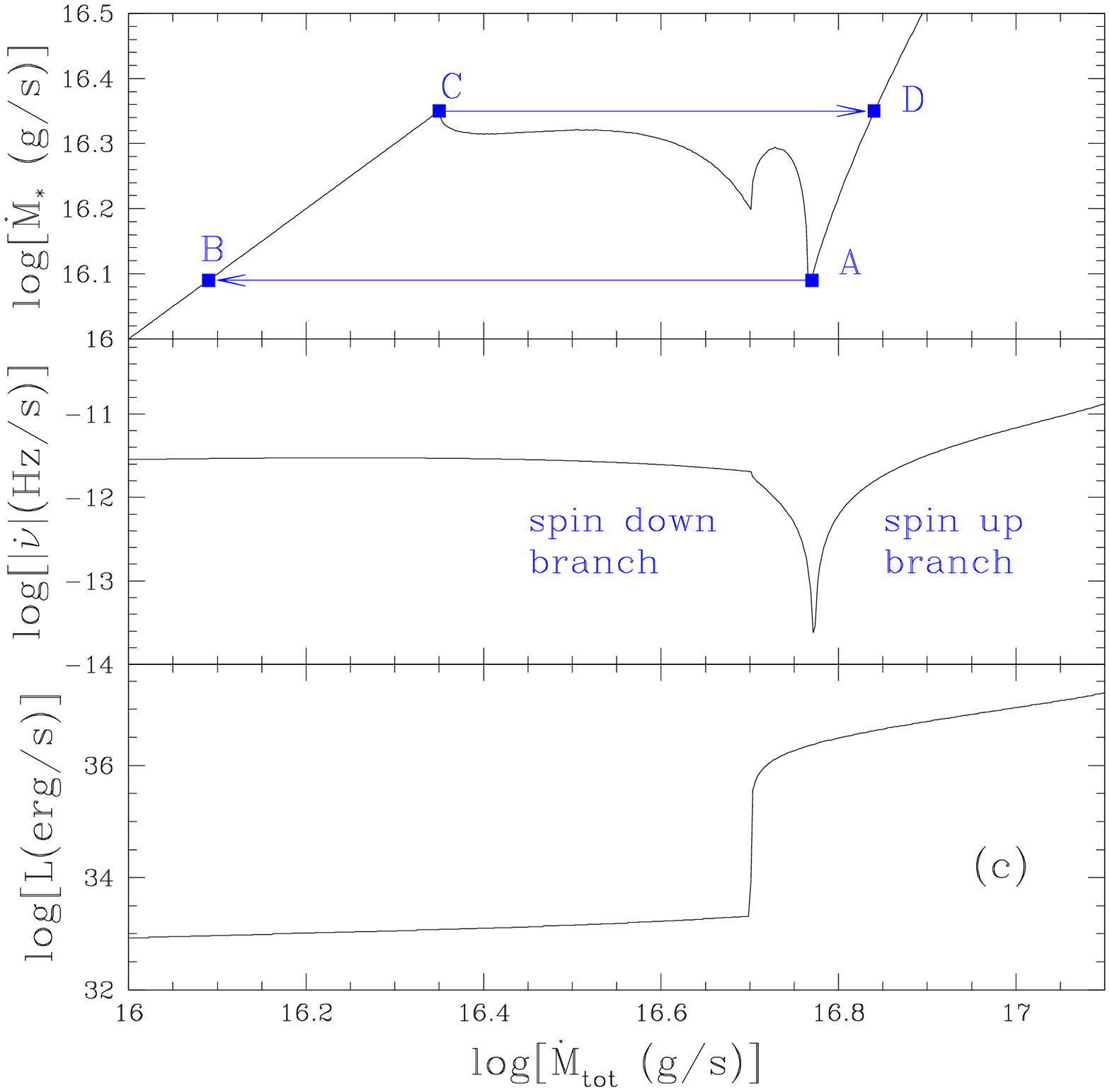}
\epsscale{1}
\caption{Various types of hysteresis limit cycles.  The system
parameters are $\nu=9$ mHz, $B=6\times 10^{13}$ G and $\beta=0$ in all
cases, while $\chi=80^\circ$ in panel (a), $\chi=50^\circ$ in panel
(b) and $\chi=47^\circ$ in panel (c).  In the top panels of each case,
the arrows indicate the points where the system ''jumps'' between
different states as a result of variations in the external mass supply
rate $\dot{M}_*$. The middle panels show that,
under most circumstances, a jump is accompanied by
a torque reversal and, in some cases, by an abrupt change
in luminosity (displayed in the bottom panels).}
\label{cycles}
\end{figure}

Next we compute the different contributions to the
luminosity. A schematic representation of these contributions
is shown in Figure \ref{lumgeom}.
Let us consider first the region of the magnetosphere in which
there is accretion ($R_M(\phi)<R_{\rm co}$). 
The accretion luminosity is given by the potential and kinetic energy
released by matter falling from the magnetospheric radius to the
surface of the neutron star; this is
\begin{equation}
L_{\rm acc}=\int_{R_M<R_{\rm co}}
\left[GM\left(\frac{1}{R_{NS}}-
\frac{1}{R_M}\right)+\frac{1}{2}\Omega^2\left(R_M^2-R_{NS}^2\right)\right]d\dot{M}_{\rm acc}\;,
\label{Lac}
\end{equation}

\begin{figure}[t]
\centering
\plottwo{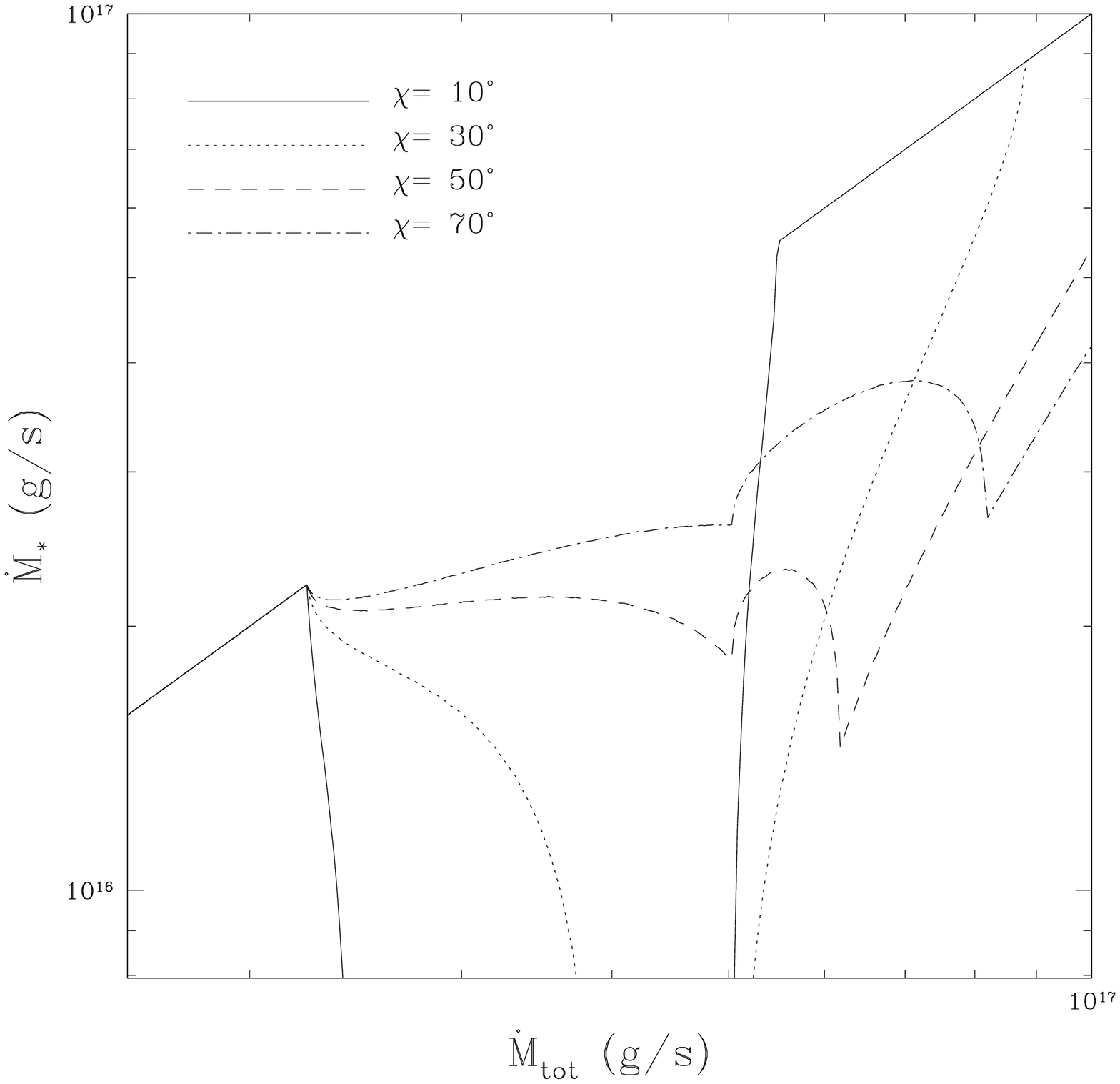}{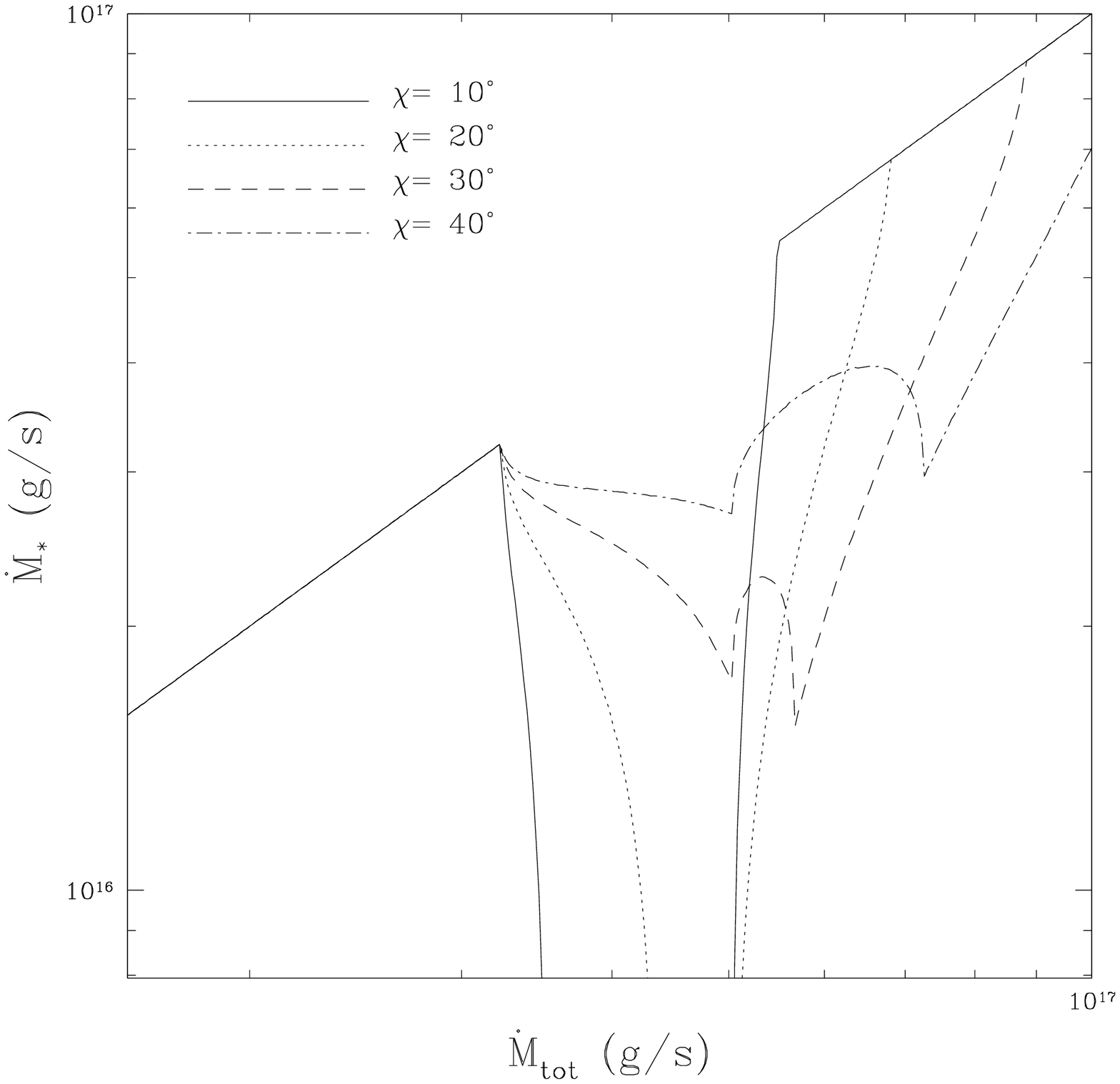}
\caption{The break of the cyclic solution in the
$\dot{M}_*-\dot{M}_{\rm tot}$ plane is shown for a system with $\nu=9$ mHz
s, $B=6\times 10^{13}$ G, $\beta=0$ (left panel) and $\beta=1$ (right
panel).  When the inclination angle is small, it is no longer possible
to find a steady state, cyclic solution. The value of $\chi$ around
which the solution breaks depends on the anelasticity parameter
$\beta$ but is independent of the values of $\nu$ and $B$.}
\label{break}
\end{figure}

where $\dot{M}_{\rm acc}$ is the fraction of
$\dot{M}_{\rm tot}$ which accretes. Next we consider the
contribution to the luminosity coming from the ''recycled matter''. This
can be calculated by summing the luminosity derived from the release of
energy of matter impacting the disk at $R_{\rm K}$, 
and the luminosity released from the same matter spiralling in the disk
from $R_{\rm K}$ back to $R_M$. This gives
\begin{equation}
L_{\rm rec}=\int_{R_{\rm co}<R_M<R_{\rm inf}}\left(\frac{v_{gen}^2}{2}-\frac{GM}{2R_M}\right)d\dot{M}_{\rm rec}
\label{lrec}
\end{equation}
where $\dot{M}_{\rm rec}$ is the rate corresponding to
the ''recycling'' part of the magnetospheric boundary
($R_{\rm co}<R_M(\phi)<R_{\rm inf}$).

\begin{figure}[t]
\centering
\plotone{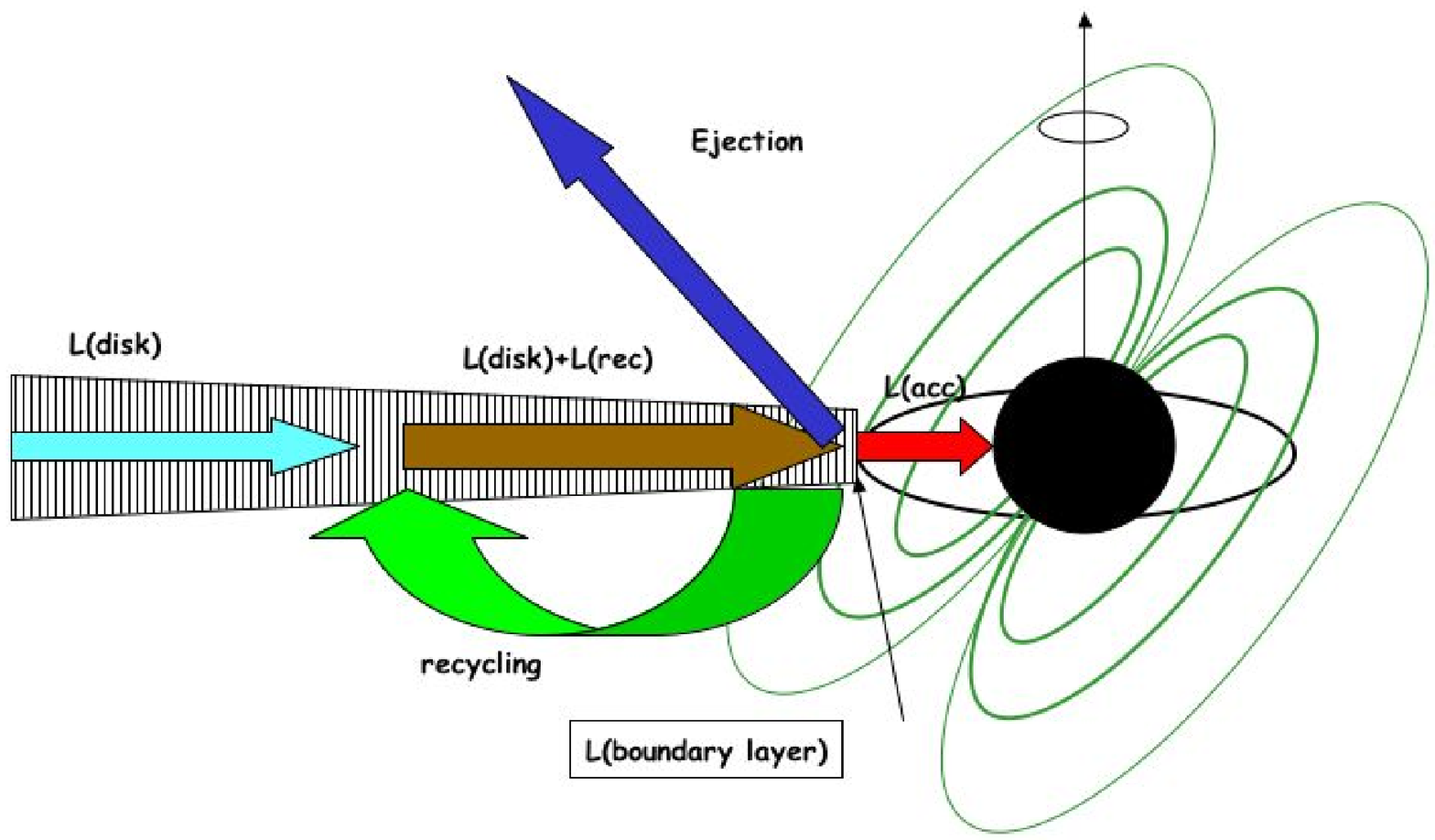}
\caption{The various contributions to the total luminosity budget for an accreting
neutron star.}
\label{lumgeom}
\end{figure}

Another contribution to the total luminosity is provided by the
release of energy in the boundary layer which separates the
magnetosphere from the Keplerian disk. This term applies to matter at
any longitude $\phi$ if we consider a completely anelastic propeller
($\beta=0$), because in this case the magnetosphere forces matter to
corotate with it during both the accretion and the propeller regime. On
the other hand, in the limit of a completely elastic propeller
($\beta=1$), this term is present only for those angles $\phi$ for which
$R_M<R_{\rm co}$ and matter is thus slowed down in
the boundary layer before it can begin falling toward the
NS. If we consider a generic value for the elasticity parameter, the
luminosity of the boundary layer can be written as:
\begin{equation}
L_{\rm BL}=\left\{ \begin{array}{ll}
 \displaystyle\frac{\dot{M}_{\rm tot}}{4\pi}\int_{0}^{2\pi}\big[R_M^2(\Omega_{\rm K}^2-\Omega_0^2)\big]d\phi
 & {\rm for}\;\; R_M<R_{\rm co} \\
 \displaystyle\frac{\dot{M}_{\rm tot}}{4\pi}(1-\beta)\int_{0}^{2\pi}\big[R_M^2(\Omega_0^2-\Omega_{\rm K}^2)\big]d\phi
 & {\rm for}\;\;R_M\geq R_{\rm co}\;.
\end{array} \right.
\label{lbl}
\end{equation} 
Finally, we have to account for the luminosity produced by 
the matter inflowing from the companion as it spirals in 
towards the magnetospheric radius in the Keplerian
disk. This contribution, which is obviously present in 
all different regimes, is given by
\begin{equation}
L_{\rm disk}=\frac{GM\dot{M}_*}{2R_M}\;.
\label{ldisk}
\end{equation} 

\begin{figure}[t]
\centering
\plottwo{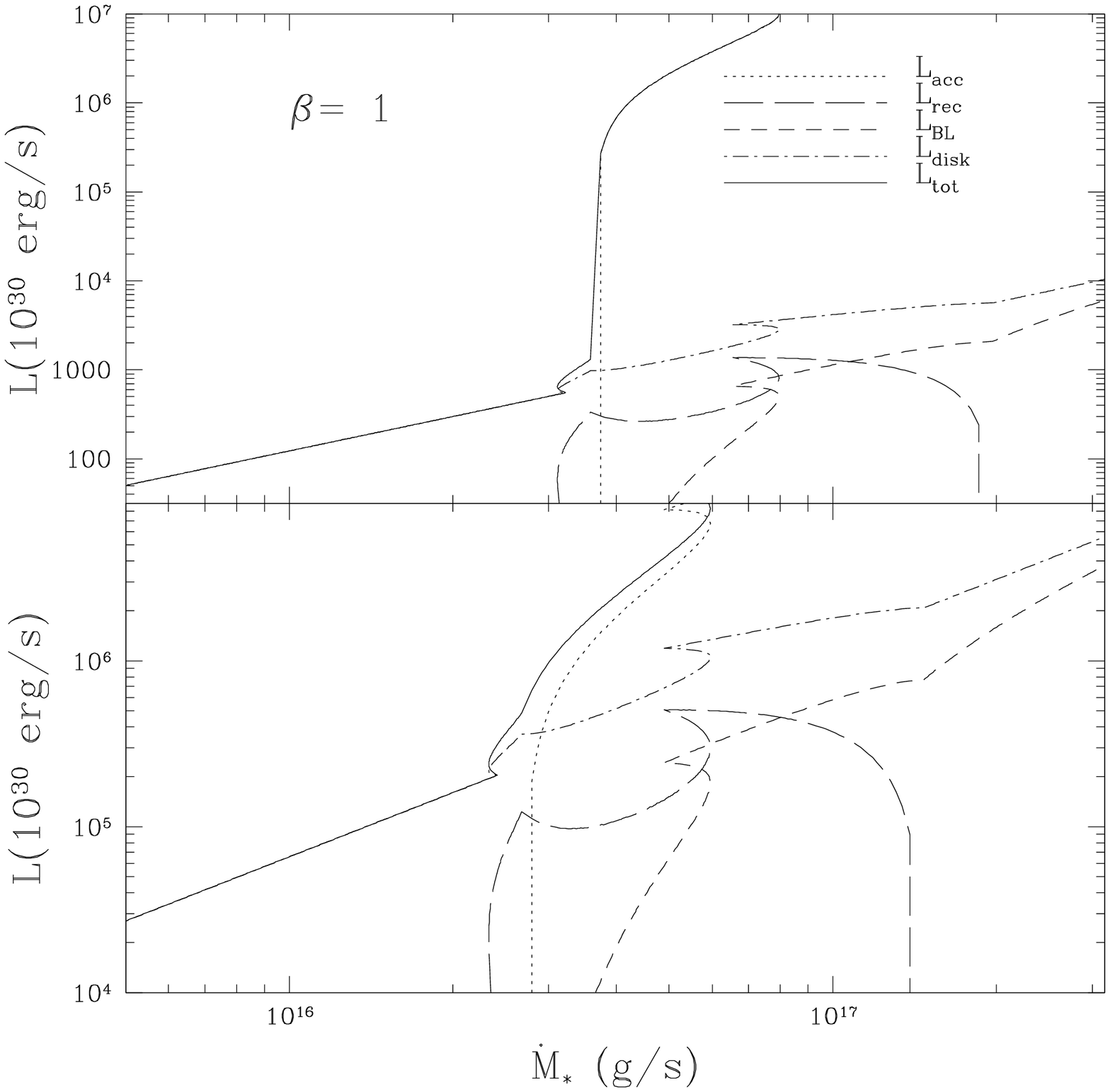}{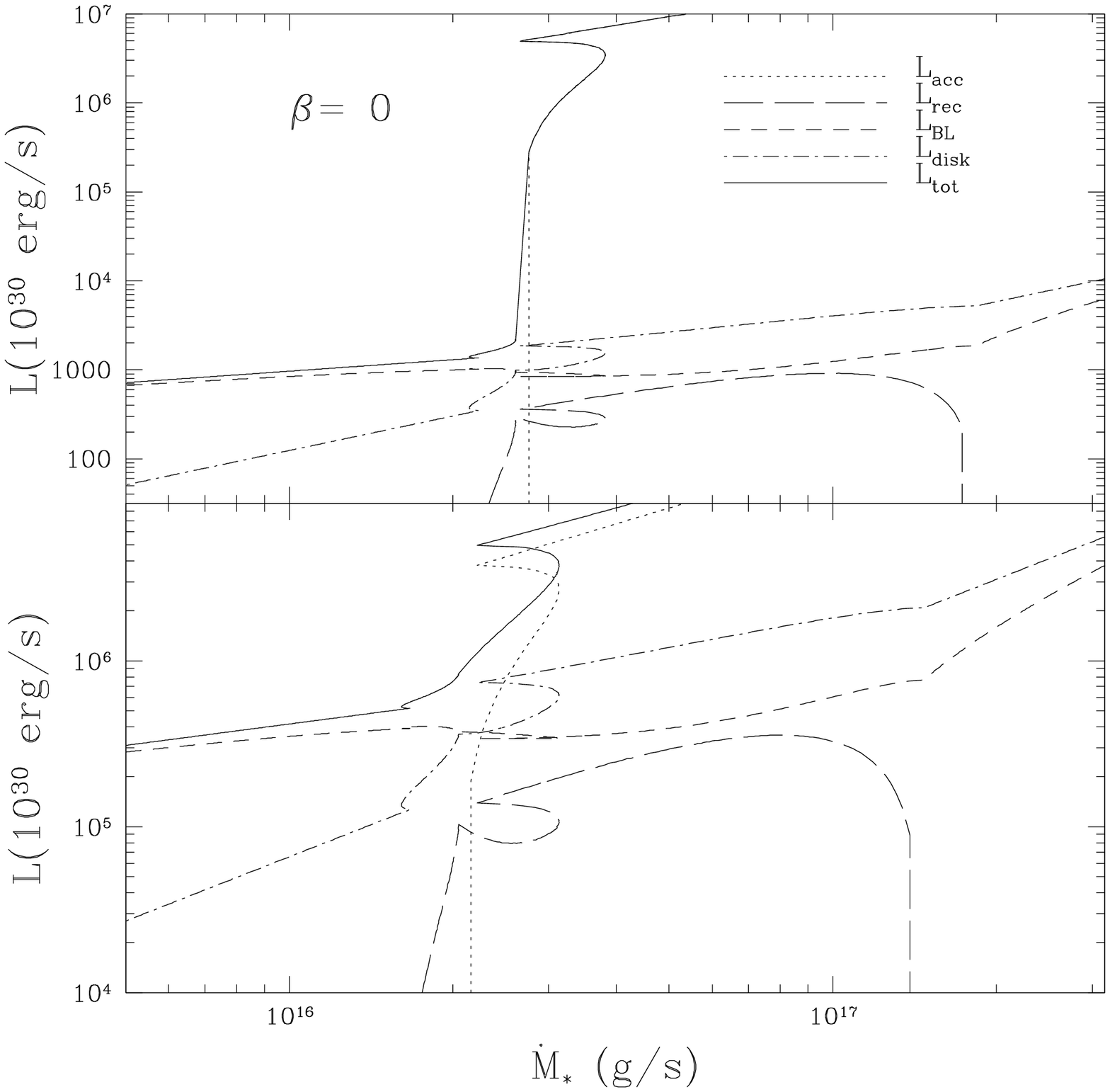}
\caption{The various contributions to the total luminosity budget
$L_{\rm tot}= L_{\rm acc}+L_{\rm rec}+L_{\rm disk}+L_{\rm BL}$ for an accreting neutron
star as function of the mass accretion rate from the companion,
$\dot{M}_{*}$. The system parameters are $\nu=9$~mHz, $B=6\times
10^{13}$~G, $\chi=80^\circ$ in the top panels and $\nu=9$~mHz,
$B=10^{9}$~G, $\chi=80^\circ$, in the bottom ones. The two limiting
cases of a completely elastic interaction (left panels) and of a 
completely anelastic interaction (right panels) are shown.}
\label{lum}
\end{figure}

It is important to emphasize that in our model both the torque and the
luminosity depend on the total mass inflow rate $\dot{M}_{\rm tot}$
at the magnetospheric boundary, and this can take different values
for the same mass accretion rate $\dot{M}_*$. 
The three panels of Figure~\ref{cycles} show the behaviour of the torque
and luminosity as a function of $\dot{M}_{\rm tot}$ for three
combinations of NS parameters.  These are chosen to represent
different types of limit cycles (also shown in the figure for
each case -- note the axes here are swapped with respect to
Figure~\ref{timeindependent} for consistency with the other
panels). In Figure~\ref{cycles}(a), a transition between the points A and B
is accompanied by a reversal from spin up to spin down, while
the jump from point C to D will cause a transition from spin down
to spin up. The luminosity is at its lowest at point B and at its
highest at point D, but the overall variation during the cycle
is well within an order of magnitude.
A more complicated cycle is depicted in Figure~\ref{cycles}(b); here a
transition from point A to B causes a spin-up to spin-down reversal,
while the opposite happens during the jump from point E to F. 
This cycle comprises also another jump, from point C to D, with both
points on the spin-down branch. The luminosity varies by
more than three orders of magnitude during the cycle, being
at its lowest during most of the spin down phase. The third
example of limit cycle, the one shown in Figure~\ref{cycles}(c), 
has only two allowed jumps, both of them
straddling the point of torque reversal, as in case a), but 
the luminosity is substantially larger when the system is on
the spin-up branch (A -- D), than when it is on the spin-down
branch (B -- C). 

Whether there exists a limit cycle depends crucially on the angle
$\chi$: this has to be large enough to ensure that some regions of the
magnetosphere are in the accretion regime while, at the same time,
others are in the propeller phase.  There exists a critical value of
the magnetic colatitude, $\chi_{\rm crit}$, below which the steady
state solution breaks into two disjoint curves and it is no longer
possible to find a cyclic behaviour through a sequence of steady-state
solutions. This is illustrated in Figure~\ref{break} for the cases
$\beta=0$ and $\beta=1$. If the accretion rate $\dot{M}_*$ from the
companion is above a certain value (which depends on $\nu$, $B$, $\chi$
and $\beta$), only one solution is available to the system, and it
corresponds to the spin up branch (see Figure \ref{cycles}). On the
other hand, if $\dot{M}_* \la \dot{M}_{\rm crit}$ (for the example
under consideration, $\dot{M}_{\rm crit}=2.4\times 10^{16}$ g/s for
$\beta=0$ and $\dot{M}_{\rm crit}=3.5\times 10^{16}$ g/s for
$\beta=0$, but it varies with $\nu$ and $B$), then multiple solutions
are available for any value of $\chi$ displayed, and the one that is
realized at any given time depends on the history of the
system. However, a cyclic jump of the solutions between the spin-up
and the spin-down branches can only be realized for angles above
$\chi_{\rm crit}$.

The critical angle ranges from about $25^\circ-30^\circ$ for $\beta=1$
to about $40^\circ-45^\circ$ for $\beta=0$, and, for a given $\beta$,
it is independent of $\nu$ and $B$.  Therefore, for the curves shown in
the figure, a limit cycle can only be achieved in the cases with
$\chi=30^\circ$ and $\chi=40^\circ$ for $\beta=1$, and in the cases
with $\chi=50^\circ$ and $\chi=70^\circ$ for $\beta=0$.  In the other
cases displayed, the $\dot{M}_*(\dot{M}_{\rm tot})$ curve is
discontinuous. The shape of the curve is such that, if the system is
spinning up, a solution on the spin-up branch can be found for any
value of $\dot{M}_*$, and therefore the system will continue spinning
up. On the other hand, if the system is originally on the spin-down
branch (which is possible only for $\dot{M}_*\la \dot{M}_{\rm crit}$),
then any decrease in $\dot{M}_*$ will keep the system on the spin-down
branch, while an increase in $\dot{M}_*$ above $\dot{M}_{\rm crit}$
will cause a jump on the spin-up branch, and from that point on the
system will be spinning up independent of the value of
$\dot{M}_*$.  Note that, depending on the angle $\chi$,
there can be spin-up solutions even at very low mass inflow rates
$\dot{M}_*$. This result is a novelty of our model, deriving from the
fact that the recycled mass component $\dot{M}_{\rm rec}$ can keep the
magnetospheric radius in ``pressure'' even if $\dot{M}_*$ is very
small.

Among all the components that make up the total luminosity, the
accretion term is the only that is certainly pulsed, since the
accretion material is funnelled by the magnetic field of the star onto
the NS magnetic poles, where its energy is released. The accretion
luminosity therefore varies with the phase of the star, resulting in a
pulsating flux.  Also the boundary layer luminosity might be pulsed at
the NS spin.  Therefore, the maximum pulsed fraction in our model is
constrained to be between $f_{\rm pul}=L_{\rm acc}/L_{\rm tot}$ and
$f_{\rm puls}= (L_{\rm acc}+L_{\rm BL})/L_{\rm tot}$.

Note that the sum of the various contributions in Equations
(\ref{Lac}), (\ref{lrec}), (\ref{lbl}), (\ref{ldisk}) can result in a
complex, non-monotonic dependence of $L_{\rm tot}$ as function of the
accretion rate from the companion, $\dot{M}_{*}$. In the classical
model of accretion onto magnetized neutron stars, the transition
between the standard accretion regime onto the NS surface to the
regime of accretion onto the magnetospheric boundary in the propeller
regime is marked by the change between the $\propto\dot{M}_*$ and the
$\propto\dot{M}_*^{9/7}$ scaling of the luminosity (Stella et
al. 1994; Campana \& Stella 2000). In the propeller phase, the
underlying assumption of these works is that the main contribution to
the luminosity derives from the disk luminosity (Eq.~\ref{ldisk}).  In
the present model, this might not be the case if there is a
non-negligible contribution to the luminosity from recycled matter.
Moreover, at low accretion rates, we find that the contribution to the
luminosity from the boundary layer generally dominates over that from
the disk for an anelastic propeller (see Figure~\ref{lum}).  For
sufficiently low values of $\dot{M}_*$ (so that $\Omega_{\rm
K}^2(R_M)/\Omega_0^2\ll 1$), the boundary layer luminosity scales as
$\propto\dot{M}_*^{3/7}$, while $L_{\rm BL}\propto\dot{M}_*^{9/7}$ at
high values of $\dot{M}_*$ (for which $\Omega_{\rm
K}^2(R_M)/\Omega_0^2\gg 1$).  When the corotation radius is of the
order of the magnetospheric radius, however, these dependences are
changed. Since the Keplerian frequency at the magnetospheric radius is
an increasing function of $\dot{M}_*$, in the propeller regime the
term $\left|\Omega_0^2- \Omega_{\rm K}^2(R_M)\right|$ decreases with
the increase of $\dot{M}_*$, while in the accretion regime the same
term increases with increasing $\dot{M}_*$.  As a result, when $R_{M}$
is of the order of $R_{\rm co}$, the luminosity of the boundary layer
has a flatter dependence on $\dot{M}_*$ for $R_M>R_{\rm co}$ and a
steeper dependence for $R_M<R_{\rm co}$.  This can be seen in
Figure~\ref{lum}. Both the case of a completely anelastic propeller
($\beta=0$), and a totally elastic one ($\beta=1$) are considered,
showing respectively the maximum and the minimum boundary layer
luminosity that the system can have. In the former case we find that, for
sufficiently low accretion rates (so that the whole magnetospheric
boundary is in the propeller regime), the boundary layer luminosity is
substantially larger than the disk luminosity.  The relative
contribution $L_{\rm BL}/L_{\rm disk}$ clearly increases as the degree
of anelasticity increases, since $L_{\rm BL}\propto (1-\beta)$.  As
the mass accretion rate $\dot{M}_*$ increases, so that at least some
regions of the magnetospheric boundary are in the accretion regime,
the disk luminosity begins to dominate over that of the boundary layer
(this is now independent of $\beta$).  However, $L_{\rm BL}$ has a
stronger dependence on $\dot{M}_*$, and, for sufficiently large
$\dot{M}_*$ that $R_M\ll R_{\rm co}$, $L_{\rm BL}$ becomes $\sim
L_{\rm disk}$.

The two panels in Figure~\ref{lum} show the cases of a slow pulsar
($\nu=9$~mHz) and a fast one ($\nu=100$~mHz). The discussion above
regarding the relative contribution of $L_{\rm BL}$ and $L_{\rm disk}$ to the
total luminosity budget holds in both cases.  Furthermore, once
accretion sets in, the accretion luminosity dominates over both
$L_{\rm disk}$ and $L_{\rm BL}$.  The slower the pulsar, the larger is this
term compared to the others. Therefore, in the accretion regime and
for $R_M\ll R_{\rm co}$, $L_{\rm tot}\propto \dot{M}_*$ as in the classical
models. However, for $R_M\sim R_{\rm co}$, the presence of the
``recycled'' term of luminosity in our model causes a non-monotonic
dependence of the total luminosity on $\dot{M}_*$, with multiple
solutions allowed.  The actual solution that is realized at any given
time will depend on the history, i.e. whether the system is on the
spin-up or spin-down branch of the limit cycle (see Figure
\ref{timeindependent}). This is an important difference of our model
with respect to the classical solution where, once the system is in
the accreting phase, the luminosity scales monotonically with
$\dot{M}_*$. On the other hand, Figure~\ref{lum} shows that there are
regions for which a small variation in $\dot{M}_*$ can cause a large
jump in luminosity. 

Similar to the luminosity, the behaviour of the torque in the
surroundings of the region with $R_{M}\sim R_{\rm co}$ is complex and
non-monotonic. Small variations in $\dot{M}$ can cause the system to
jump between states with opposite sign of the torque.  Within this
region, because of the complex dependence of both $L$ and
$\dot{\Omega}$ on $\dot{M}_*$, our model does not make any specific
prediction regarding correlations between torque and luminosity.  In
most situations, these are expected to be uncorrelated, and different
types of limit cycles (see Figure~\ref{cycles}) will generally lead to
different behaviours in the various spin-up and spin-down phases.

\begin{figure}[t]
\centering 
\plotone{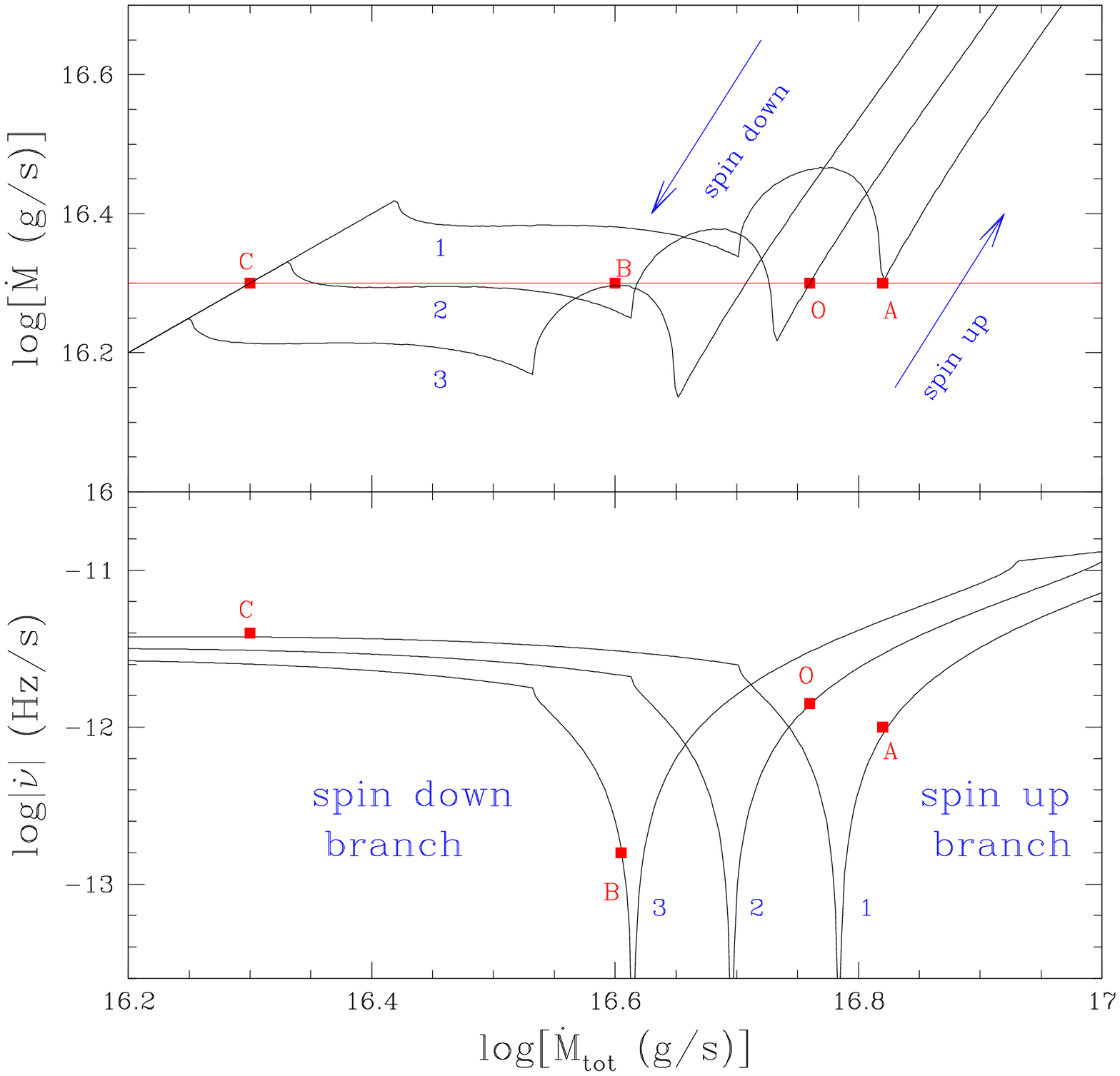}
\caption{Variation of the function $\dot{M}(\dot{M}_{\rm tot})
\equiv \dot{M}_{\rm acc} + \dot{M}_{\rm eje}$ (top panel)
and of the corresponding spin rate variation (bottom panel)
at three different times during a spin up/spin-down cycle. 
The system parameters are $B=6\times 10^{13}$ G, $\chi=45^\circ$
and $\beta=0.3$. The three curves correspond to frequencies
$\nu=9.1$~mHz (curve 1), $\nu=8.3$~mHz (curve 2) and $\nu=7.7$~mHz (curve 3). 
Allowed states for the system are only those satisfying the condition
$\dot{M}_*= \dot{M}(\dot{M}_{\rm tot})$; in this case, $\log(\dot{M}_*)=16.3$.
When multiple solutions are allowed,
the state in which the system will be found depends on its previous history.}
\label{proptime}
\end{figure}

\subsection{Cyclic spin-up/spin-down evolution at a constant $\dot{M}_*$}

In systems in which mass transfer takes place through Roche Lobe
overflow, the rate at which material is fed to the disk is expected to
be roughly constant, or characterized by relatively low-amplitude,
long-term variations. We are not concerned in this section with the
accretion disk instabilities that likely give rise to the very large
amplitude variations of the mass inflow rate in binary X-ray transient
systems. Rather, in the following we describe how recurrent episodes of
spin up and spin down can be achieved in our model in response to a
strictly constant accretion rate $\dot{M}_*$ from the companion star.

Figure~\ref{proptime} shows the behaviour of the curve $\dot{M}=
\dot{M}_{\rm eje} + \dot{M}_{\rm acc}$ (top panel) and the
corresponding frequency derivative, $\dot{\nu}$, (bottom
panel) as a function of $\dot{M}_{\rm tot}$ and for different values
of the period (corresponding to different times).  The parameters $B$,
$\chi$ and $\beta$ are the same in all cases.  They yield a limit
cycle of the type described in Figure \ref{cycles}(b).  While the
specific points of torque reversal will vary depending on the type of
cycle (as shown in the various examples of Figure \ref{cycles}), the
underlying structure determining the transitions is the same in all
cases and therefore we analyze in detail only one of the possible
scenarios.

In order to illustrate how the spin up/spin down states are achieved
at a constant $\dot{M}_*$, let's start, say, with the system at a
frequency $\nu$ so that the corresponding $\dot{M}(\dot{M}_{\rm tot})$
curve is the one labeled ``2'' in Figure~\ref{proptime}, and let's
assume that the system is in a spin-up state. The intersection between
the curves $\dot{M}$ and $\dot{M_*}$ on the spin-up branch of the
cuspid determines the value of $\dot{M}_{\rm tot}$, $\dot{M}_{\rm
tot,sol}$, corresponding to the allowed spin-up state for that value
of the frequency.  This value of $\dot{M}_{\rm tot,sol}$ in turn
determines the value of the frequency derivative at that point in time
(point ``O'' in both panels of the figure).  The frequency
at time $t+dt$ is simply determined as
$\nu(t+dt) = \nu(t) + d\nu(\dot{M}_{\rm
tot,sol})/dt$. As the pulsar spins up, the curve ``2'' moves towards
curve ``1'' until the point at which the spin-up branch of the
solution rises higher than the system $\dot{M}_*$ (point A). From that
point on, the only possible state for the system that satisfies the
condition $\dot{M}=\dot{M}_*$ is the one corresponding to point C in
the figure, on the spin-down branch (negative torque). Once again, the
new (current) value of $\dot{M}_{\rm tot, sol}$ determines the actual
value of $\dot{\nu}$ (corresponding point C in bottom panel)
which is used for the next time step to determine the new $\nu$.
While on the spin down branch of the solution, the curve $\dot{M}$ now
moves from the curve ``1'' towards the curve ``2'' and then
``3''. Spin down continues until this branch of the solution does not
intersect any longer the $\dot{M}_*$ line (point B), at which point
the only allowed state for the system to be is on the spin-up branch,
and the system reverses from spin down to spin up. This is the
beginning of a new cycle.

\begin{figure}[t]
\centering
\plotone{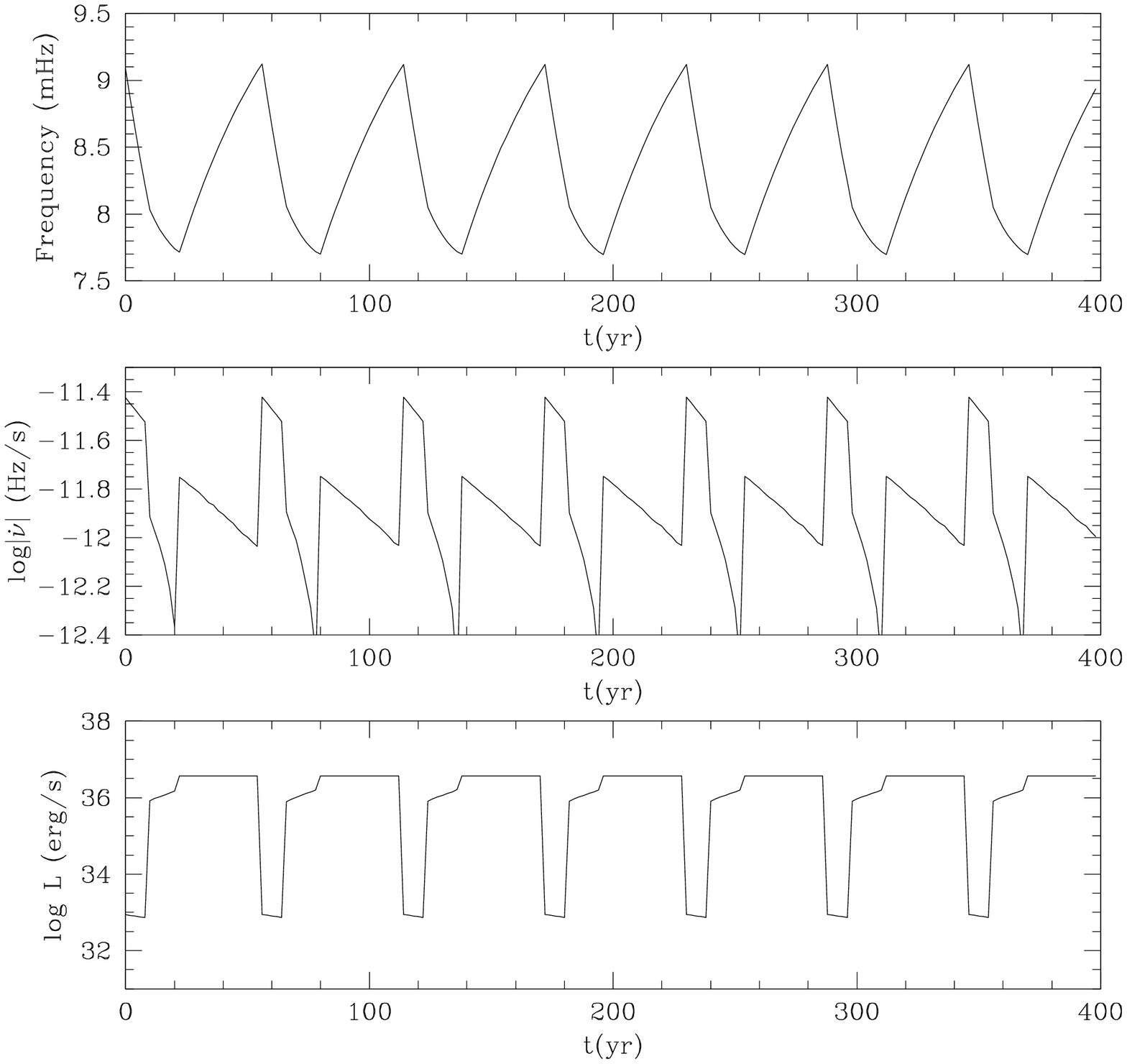}
\caption{An oblique NS rotator with magnetic field $B=6\times 10^{13}$ G,
inclination angle $\chi=45$ deg, and elasticity parameter $\beta=0.3$
is able to reproduce the main spin-up/spin down characteristics of GX 1+4.
The luminosity is comparable during the spin-up and spin-down phases,
except for a few years at the beginning of the spin-down phase, when it
drops abruptly.}
\label{gx}
\end{figure}

In this model, the points of spin reversals are determined by the
maximum and minimum of the $\dot{M}(\nu)$ curve. The shape of this
curve depends on the anelastic parameter $\beta$ and on the angle
$\chi$. For a given $\beta$ and $\chi$, a change in the strength of
the magnetic field simply results in a shift of the curve without a
change in shape: a higher $B$ field would move the curve to higher
values of $\dot{M}_{\rm tot}$, therefore resulting in stronger spin-up
and spin-down torques, and hence in a shorter timescale for torque reversals.
For the model to work as described, it is clear that the
points where the solution jumps must straddle the point of torque
reversal. We find this to be the case for a wide range of combinations
of $\chi$ and $\beta$. However, for each value of $\beta$, there is a
narrow range of angles $\chi$ for which the torque inversion point
falls outside the allowed region for the transitions. For these
particular and rare cases, the system would tend towards the point
$\dot{\Omega}_0=0$ and remain there, for a strictly constant
$\dot{M}_*$.  However, small fluctuations in $\dot{M}_*$ can still
cause the system to jump from one solution to another.  For the rest
of this discussion we will focus on the greatest majority of cases for
which torque reversals naturally occur at $\dot{M}_*$ = const, unless
we explicitly state otherwise.

If the magnetic colatitude angle $\chi$ is larger than $\chi_{\rm crit}$, the
system is bound to end up in a cyclic sequence of spin-up/spin-down
transitions. In fact, as it can be seen from Figure \ref{proptime}, if
the system starts with a much larger frequency than the maximum
frequency in the cycle, $\nu_{\rm max}$, it will spin down since only one
solution (on the spin down branch) is allowed as long as $\nu>\nu_{\rm
max}$.  Similarly, if the system starts with a frequency much smaller
than the mimimum frequency in the cycle, $\nu_{\rm min}$, it will spin
up as only one solution (on the spin up branch) is allowed as long as
$\nu<\nu_{\rm min}$.  Therefore, our model predicts that the system,
independent of the initial conditions, eventually settles in a
region where there are cyclic transitions between spin-up and spin-down 
states. This limit cycle is not induced by external perturbations,
but is the natural equilibium state torwards which the system tends.

\section{Application of our model to persistent X-ray pulsars}

In the following, we will apply our model to two objects for which
long term-monitoring showed a marked transition between a spin-up and
spin-down phase. We will then discuss the way our model can be
generalized to other cases where short-term episodes of
spin-up/spin-down are superimposed onto longer term spin-up or
spin-down trends.  The most comphrensive monitoring of the spin
behaviour of accreting X-ray pulsars in binaries is given in Bildsten
et al. (1997), and here we briefly summarize the observations for the
two cases that we model.

\subsection{GX 1+4}
 
GX 1+4, discovered in 1970 through an X-ray balloon experiment (Lewin,
Ricker \& McClintock 1971) is an accreting X-ray pulsar binary hosting an 
M red giant (Davidsen et al. 1977); the orbital period is likely to
be of a few years (Chakrabarty \& Roche 1997). Early observations
through the 1970s showed that this source was spinning up at a very
high pace with a spin-up timescale $|\nu/\dot{\nu}|\sim 40$ yr. The
frequency changed from $\sim 7.5$ mHz to $\sim 9$ mHz during the first
15 years of observations.  In the early 1980s, however, the flux
dropped abruptly and the source could not be detected by {\em
Ginga}. Given the sensitivity of the instrument, the flux must have
decresead by more than two orders of magnitudes for a few years.  Once
its flux raised, the source could be monitored again, and it was found
to spin down on a timescale comparable to the previous spin-up
timescale (Makishima et al. 1988).

A solution that closely reproduces the observed source behaviour was
found by running the time-dependent code described in \S 2.4 for a
range of parameters $B, \chi,\beta$. The corresponding value of $M_*$
is determined so that the point of torque reversal of the system
between spin up and spin down matches the observed value. The larger
the magnetic field, the larger $\dot{M}_*$ and hence the torque,
and therefore the more rapid the timescales of the torque transitions
are; the parameters $\chi$ and $\beta$, by determining the shape of
the $\dot{M}$ curve, especially influence the total frequency
range $\nu_{\rm max}-\nu_{\rm min}$ that the system spans in a cycle.
 
For the case of GX~1+4, we found that a good choice of parameters is
the combination $B=6\times 10^{13}$ G, $\chi=45^\circ$, and
$\beta=0.3$.  These yield a cycle of the type displayed in 
panel (b) of Figure~\ref{cycles} and in Figure~\ref{proptime}.
In particular, the parameters $B$, $\chi$ and $\beta$ used
in Figure~\ref{proptime} are the same as those used for GX~1+4.
The accretion rate provided by the donor companion must
be $\dot{M}_*\sim 2\times 10^{16}$ g/s in order to produce a turnover in
frequency around 9 mHz.  With this choice of parameters, Figure~\ref{gx} 
shows the behaviour of the system that our model predicts.
Cycles of spin-up/spin-down alternate in response to torque
reversals. The luminosity of the source is comparable 
during the spin-up and spin-down phases, except 
for a few years around the time of spin reversal from spin
up to spin down, when it drops abruptly. This is
due to the fact that, after the system has ``jumped'' to point ``C''
in Figure~\ref{proptime} (at the beginning of the spin down phase),
there are no regions in the magnetosphere-disk boundary where
accretion onto the NS can take place, the NS is in the propeller
regime, and therefore $\dot{M}_{\rm acc}=0$ (see Figure~\ref{mdot}).
During that time, the only contribution to the luminosity comes from
the disk and the boundary layer, which are however much smaller than
the accretion luminosity (since this is a slow pulsar).  A prediction
of our model is that, while large drops in luminosity can be expected
when the system reverses from spin up to spin down, they should not
occur in correspondence of the spin-down/spin-up transition, because
when this transition occurs (refer to the jump from point ``B'' to
point ``A'' in Figure \ref{proptime}), most regions at the
magnetospheric boundary are allowed to accrete.
While these overall features are generally robust predictions of our model,
the detailed variation of $\dot{M}$ (and hence the luminosity) with torque 
shown in our examples should not be taken too rigorously. These variations 
depend on the shape of the $\dot{M}(\dot{M}_{\rm tot})$ curve, and this is in turn
determined by the shape of the magnetospheric boundary as a function of time.
As discussed in \S 3.3, a number of effects neglected here can influence this shape,
and hence affect the detailed behaviour of the solution.
In particular, note that observations of GX 1+4  show that luminosity and torque
strength are correlated during part of the spin down phase (Chakrabarty et al. 1997b). 
This feature is not reproduced by the current version of our model.

\subsection{4U 1626-67}

4U 1626-67, discovered by SAS-3 in 1977 (Rappaport et al. 1977) is an
ultracompact binary with an extremely low-mass companion (Levine et
al. 1988; Chakrabarty et al.  1997a) and a 42 minute orbital period
(Middleditch et al. 1981).  During the first $\sim 20$ years of
observations, the source was found to spin up with a timescale
$\nu/\dot{\nu}$ of about 5000 yr. The frequency increased from 130.2
mHz to about 130.5 mHz, at which point the source started to spin
down. Unlike the case of GX 1+4, there was no evidence for a large
change in the bolometric luminosity of the source during the
transition.

\begin{figure}[t]
\centering
\plotone{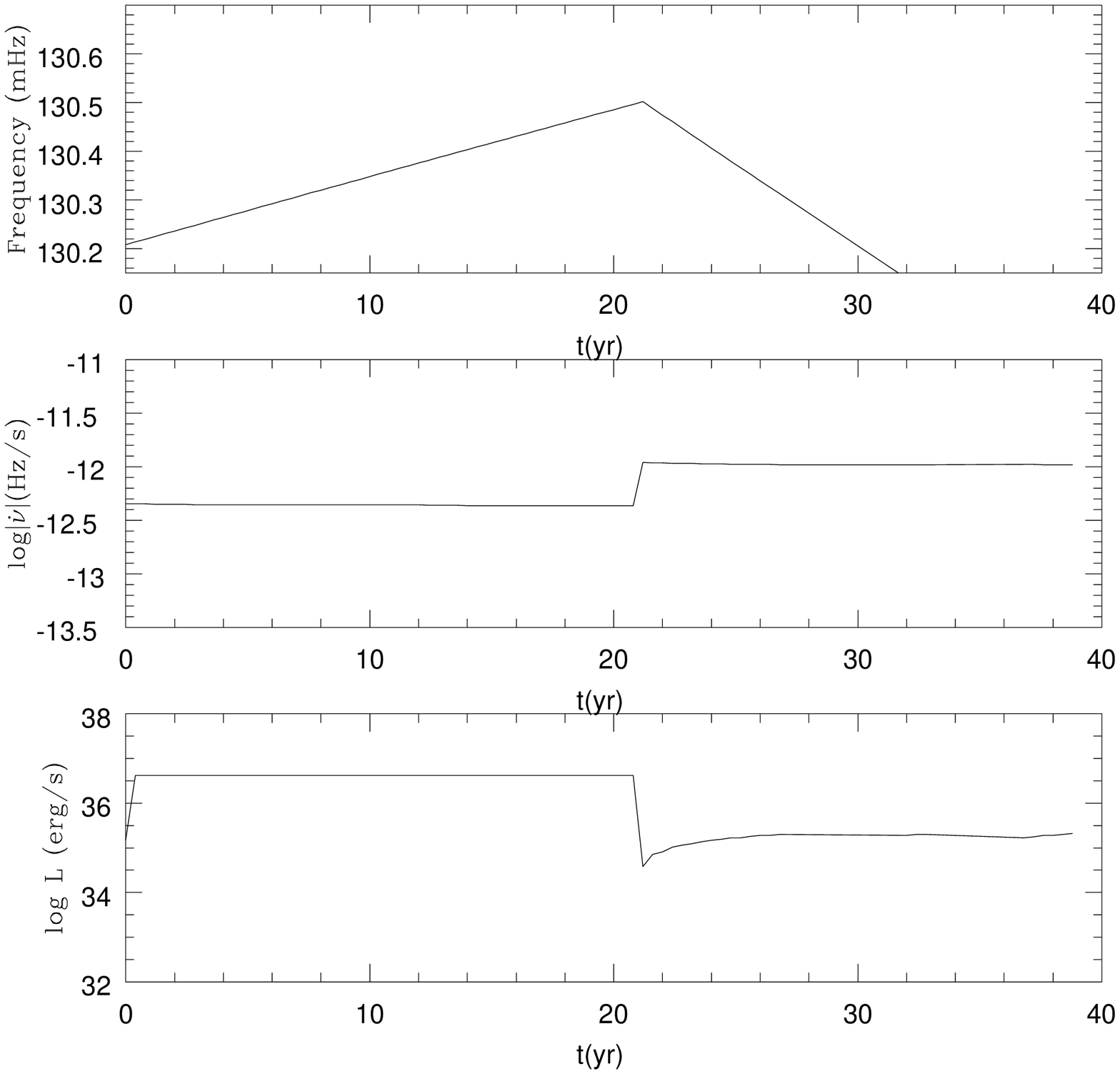}
\caption{An oblique NS rotator with magnetic field $B=2.5\times 10^{12}$ G,
inclination angle $\chi=68^\circ$, and elasticity parameter $\beta=0$
is able to reproduce the main spin-up/spin down characteristics of 4U 1626.}
\label{4u}
\end{figure}

The very long timescale for spin reversal of this source (due to a
smaller torque compared to the case of GX~1+4) requires a smaller
magnetic field. We found that our model yields a reasonable match to
the observations with the choice of parameters $B=2.5\times 10^{12}$
G, $\chi=68^\circ$, $\beta=0$. The corresponding solution found with
our model is displayed in Figure~\ref{4u}. The upper panel shows only
one spin-up/spin-down torque reversal, since the complete
spin-up/spin-down cycle, of the order of several thousand years, lasts
much longer than the observed time.  Although the luminosity somewhat
drops around the time of spin reversal, it does so to a lesser extent
and for a much shorter time than for the case of GX~1+4.  The reason
for these differences lies in the variation of the shape of the
function $\dot{M}(\dot{M}_{\rm tot})$ for different choices of the
parameters $\chi$ and $\beta$. The parameters that best match the
solution for 4U 1626-67 yield a cycle of the type in panel (a) of
Figure~\ref{cycles}. The transition from spin up to spin down (point A
to B in the figure) is accompanied by a less dramatic variation in
luminosity than it is for the cycles of the type shown in panels (b)
and (c). Note how, for this source, since the observation window is
much smaller than the timescale for torque reversal, other torque
inversions are not expected in the near future, unless induced by
external perturbations.

\subsection{Generalizations and
limitations of our model}

The two examples given above, for two sources spinning up and down
at very different rates, show that our model can reproduce different
types of cyclic behaviours.  In the two cases discussed, we
assumed that the mass accretion rate from the companion, $\dot{M}_*$,
does not vary with time. Under this assumption, our model predicts
that the points of torque reversals will always occur at the same
value of the frequency. On the other hand, if the donor accretion rate
varies with time, this will no longer be the case. If $\dot{M}_*$
increases with time, then the points of torque reversals will occur at
larger frequencies as time goes on. Viceversa if $\dot{M}_*$ decreases
with time, then the points of torque reversals will occur at smaller
and smaller frequencies with time. A combination of discrete 
states in an oblique rotator (producing cyclic torque reversals), with
longer-term variation in the external $\dot{M}_*$ can produce a
long-term spin evolution with superimposed shorter cyclic episodes of
spin up and spin down. 

Also note that, depending on the system parameters (namely the
inclination angle $\chi$ and the elasticity parameter $\beta$, which
determine the shape of the $\dot{M}(\dot{M}_{\rm tot})$ curve, and
hence the points of torque reversals), the transition from a state of
spin up to a state of spin down can result in a period of time during
which accretion is completely inhibited (i.e. $\dot{M}_{\rm acc}=0$)
and the luminosity is orders of magnitude lower (unless the pulsar has
a very fast spin in the ms range and the luminosity of the disk and
the boundary layer are conspicuous even when $\dot{M}_{\rm acc}=0$).
The system can then behave as a ``transient'' even when the accretion
rate from the companion is constant. 

In the present (simplest) version of our model,  the frequency range
$(\nu_{\rm max}-\nu_{\rm min})$ spanned in a spin-up/down cycle cannot
however be made arbitrarily small. In order for the torque reversals to occur
at constant $\dot{M}_*$ and without any other external perturbation, 
the curves $\dot{M}(\nu_{\rm max})$ and $\dot{M}(\nu_{\rm min})$
(curves 1 and 3 respectively in Figure~\ref{proptime})
must be such that the two points of torque reversals (A and B
in Figure~\ref{proptime}) satisfy the conditions 
$\dot{M}_{\rm A}(\nu_{\rm max})=\dot{M}_*$ and 
$\dot{M}_{\rm B}(\nu_{\rm min})=\dot{M}_*$ respectively. Arbitrarily small
cycles require arbitrarily small loops in the $\dot{M}(\dot{M}_{\rm tot})$
curve, so that the inversion points can be extremely close. 
This cannot be achieved with the current version of our model, 
in which the shape of the $\dot{M}(\dot{M}_{\rm tot})$ curve
(and hence the ``size'' of the loop around the points
of torque reversals) depends only on the inclination
angle $\chi$ and the elasticity parameter $\beta$.  However,
there are a number of effects that we have neglected here,
and which could be potentially important for small-scale torque reversals.
In particular, if the disk plane is not orthogonal to
the NS rotation axis, a precession of the disk around the spin axis
can be induced (Lai 1999), producing a time-dependent modulation
of the various regimes on a time scale on the order of the spin period
of the star. We reserve to future work a more comprehensive 
exploration of the physical effects that influence the magnitude
and frequency of the torque reversals. 

\section{Summary and Discussion}

A magnetic rotating neutron star surrounded by an accretion disk is an
intuitive example of an accreting system in which the conditions can
be realized such that a fraction of the matter is accreted, another
fraction is ejected and completely unbound from the system, and
another part is propelled out but does not possess enough energy to
unbind, and therefore falls back onto the disk, getting
``recycled''. We have shown that, for a given mass rate supply from
the companion, accretion with the mass feedback term included leads to
multiple available states for the system, characterized by different
(and discrete) values of the total mass inflow at the magnetospheric
boundary.  The luminosity in each of these states is generally
different, as it depends on the relative amounts of the various
components of the total mass inflow rate.  The available states often
straddle the point of torque reversal, and therefore correspond to
states with opposite sign of the torque.

The character of the solutions is essentially determined by the
inclination angle $\chi$ of the NS axis with respect to the disk.  At
angles $\chi\la \chi_{\rm crit}$, the limit cycle breaks down. In this
case, for an external mass supply larger than a critical value (which
depends on the system parameters), the system can only be on the spin
up branch. For accretion rates smaller than this critical value, both
the spin up and the spin down branches of the solution are possible,
and the one that is realized will depend on the history of the
system. After a sufficiently long time, however, if the system is
spinning down, the available solutions will be drifting and the source
will jump out of the spin-down branch, and continue evolving on the
spin-up branch.  For $\chi\ga \chi_{\rm crit}$, cyclic transitions
between states of opposite torque can be realized even at a constant
value of the accretion rate from the companion. This is a particularly
nice feature of our model: {\em periodic variations between spin up and
spin down states take place without requiring the presence of any
external, periodic, and fine-tuned perturbation.} Most importantly, we
have shown that periodic, cyclic episodes of spin up/spin down
behaviour {\em must} be realized in a number of situations. While in
the classical theory of accreting X-ray binaries (where the effect of
mass feedback is not accounted for) the system is expected to
eventually settle at the equilibrium frequency which matches the
Keplerian frequency at the magnetospheric boundary, in our model,
where recycling is accounted for, the system will eventually settle
around a limit-cycle behaviour in which different spin derivative and
luminosity states alternate, recurrently.  The points of spin reversal
and the timescales of the torque reversals
depend on a combination of factors, namely the accretion rate from the
companion, the magnetic field of the NS, the inclination angle of the
NS axis, and the degree of anelasticity at the disk-magnetospheric
boundary.

In the case of the two X-ray binaries GX~1+4 and 4U~1626-67, we have
determined a set of parameters $B,\chi,\beta$ which is able to
reproduce the main features of their timing behaviours, such as the
timescales and frequency span of the transitions, as well as the large
luminosity drop observed around the transition spin up-down in the
case of GX~1+4 but not of 4U~1626-67. The correlation between torque
strength and luminosity in the spin down phase observed in
GX~1+4 (Chakrabarty et al. 1997b) is however not reproduced by the
present scenario.  On the other hand, we still need to emphasize that
ours is a very simplified model and therefore the detailed behaviour
of our solution should not be considered too rigorously: while our
model appropriately accounts for the material torque at the
disk-magnetospheric boundary when a fraction of mass is recycled, it
neglects other possible sources of torque, such as magnetic stresses
(e.g. GL) or magnetically driven outflows in an extended boundary
layer (Arons et al. 1984; Lovelace et al. 1995). The presence of other
torque terms could modify the character of the solutions if
non-material torques dominate over the material one. A general
treatment that includes all possible sources of torques is beyond the
scope of this paper, especially since the relative strength of the
various terms would be hard to estimate from first principles.

Finally, while the details of the solutions that we have discussed
specifically apply to the case of a rotating neutron star accreting
from a disk fueled by a companion star, the general feature of a
multeplicity of states available for a given mass inflow rate of matter
can probably be generalized to other
accreting systems in which ``recycling'' occurs. An example is that
of an accretion disk around a rotating black hole. Numerical
simulations (e.g. Krolik et al. 2005) show that, while a fraction of
the accreting mass is ejected through a jet, another fraction, of
slower velocity and at larger angles from the jet axis, falls back
into the disk, getting recycled. It would be interesting to include
this mass feedback process into numerical simulations of accretion
disks around black holes, and investigate whether the discountinuos
states and cyclic behaviour might ensue in those cases as well.

\acknowledgments We thank an anonymous referee for useful and constructive
comments which improved the presentation of our paper. 
RP thanks the Department of Astrophysical Sciences at
Princeton University for its kind hospitality and financial support
during the time that most of this work was carried out.
LS ackowledges useful discussions with L. Burderi, T. Di Salvo and
M. Vietri in the early stages of this work. This work was partially
supported through a MIUR-PRIN grant. 

\newpage

\centerline{APPENDIX}

\appendix 

Here we justify our assumption that, during the rotation of the
magnetosphere, matter in the disk is able to fill the region
that separates the disk and the magnetospheric flow on a timescale
shorter than (or comparable to) the spin period of the star.
  
Let $\tau_{\nu}=R^2/\nu=R/v_R$ be the viscous timescale in the disk,
where $R$ is the radial distance from the star, $\nu$ the kinematic
viscosity coefficient and $v_R$ the radial velocity in the disk.  In
the reference frame of the disk (in which $\tau_\nu$ is measured),
the stellar rotation time is
$\tau_{\rm rot}=2\pi/\vert\Omega_0-\Omega_{\rm K}\vert$.  Using the thin disk
approximation, the disk height $H$ can be written as $H=fR$, where
$f<<1$ is a numerical factor that can be assumed approssimatively
costant for small variations of the radial distance from the NS
(typically $f\sim1/10$). Furthermore, using the 
$\alpha$ prescription for the viscosity (Shakura \& Sunyaev 1973), we can write
$v_R=\alpha{v_s^2}/{v_{\rm K}}=\alpha f^2 v_{\rm K}$
where $v_s$ is the sound speed in the disk and 
we have assumed $v_s/v_{\rm K}\simeq H/R$. 

Let us consider first the propeller regime ($R_M>R_{\rm co}$). 
For $R=R_M$, we obtain that 
$\tau_{\nu}<\tau_{\rm rot}$ only if
\begin{equation}
R_{\rm co}<R_M<(1+2\pi\alpha f^2)^{2/3} R_{\rm co}\;
\label{condition}
\end{equation}
which corresponds to a very narrow region around the corotation
radius. Beyond this region, the viscous timescale becomes too long
to permit a replenishment of the inner regions of the
disk as the star rotates. Here another mechanism is needed to
justify our assumption. Indeed, in the propeller regime, 
the surface of separation between the
magnetospheric and disk flow is Kelvin-Helmholtz (KHI) unstable due to
the large shear velocity (Wang \& Robertson 1985; Spruit \& Taam
1993). In the frame corotating with the NS this velocity
is $v_{\rm rel}=R_M [\Omega_*-\Omega_{\rm K}(R_M)]$.
Because of the KHI, matter in the disk is mixed with the
NS magnetic field lines, thus mantaining a strong interaction
between the disk and the magnetosphere. 

The characteristic timescale for the development of the KHI (in the
direction of the shear motion) can be estimated as (e.g. Stella \& Rosner
1984) $\tau_{KH}\approx 4\pi (k \vert v_{\rm rel}\vert)^{-1}$, where $k=2\pi/\lambda$
is the wave vector of the perturbation which inizializes the
instability.  The condition that the KHI develops within a time
shorter than the local timescale $\tau_{\rm rot}/2$ is hence satisfied for
wave vectors $k>2/R$.  Furthermore, in order for the interaction
between the disk and the magnetospheric flow to be maintained
throughout the rotation of the star, the KHI must be able to mix disk
matter and magnetic field lines at least on a distance $d\sim [R_M(0)
-R_M(\pi/2)]\la 0.5 R_M(0)$ (see Eq.(\ref{mrad}) and Fig.1).
The simulations of Wang \& Robertson (1985) show that perturbations
of lengthscale $\lambda$ become rapidly unstable and evolve into
elongated vortices of magnitude comparable to $\lambda$. 
This means that a perturbation of length $\lambda$ is able to produce
mixing between matter and field lines on a distance scale of the same order.
Wang \& Robertson also argue that the dominant
mode of the instability will likely be the one just sufficient to offset the
effect of viscous damping through the turbulent motions in the shear
layer. In our case this condition traslates into
$\lambda/2\pi v_t\sim(\Omega_*-\Omega_{\rm K})^{-1}$
where $v_t$ is the turbulent velocity. If we choose $v_t\sim v_s$ and
use $f\sim H/R$, we can roughly estimate
\begin{equation}
\lambda\sim\frac{2\pi f}{(\frac{R_M}{R_{\rm co}})^{3/2}-1}R_M\;,
\label{lam}
\end{equation}
that is of the order required to cover the radial extension $d$
discussed above (here we have used the fact that in our model the
propeller regime typically occurs for $R_M$ within the range 
$R_{\rm co}<R_M<(1.6-1.7)R_{\rm co}$).  
Since the wavelength in Eq(\ref{lam}) satisfies the condition $k>2/R_M$,
the dominant mode of instability
develops in a shorter time than the local dynamical
timescale, and therefore the KHI is able to
maintain a close interaction between the disk and the magnetosphere
on this timescale.

Let us now consider the accretion regime ($R_M<R_{\rm co}$). Using the same
derivation as above, the analogous of Eq.(\ref{condition}) is
\begin{equation}
(1-2\pi\alpha f^2)^{2/3} R_{\rm co}<R_M<R_{\rm co}
\label{condition2}
\end{equation}
which is again a narrow region in the vicinity of the corotation
radius. In the accretion regime however, considering the argument used in the propeller case
(where now $v_{\rm rel}=R_M (\Omega_{\rm K}-\Omega_0)$), we obtain the same
conclusion about the efficiency of the KHI, and the analogous of equation \ref{lam} is now
 \begin{equation}
\lambda\sim\frac{2\pi f}{(1-\frac{R_M}{R_{\rm co}})^{3/2}}R_M\;,
\end{equation} 
which clearly satisfies the requirement $ \lambda\gtrsim 0.5 R_M$ for
any value of $R_M$ in the region of interest.  Furthermore, after the
KHI has brought matter just inside the magnetospheric radius, the
enhanced contribution of the gravitational with respect to the
centrifugal force, forces matter to fall toward the NS also under the
effect of the Rayleigh-Taylor instability (Arons \& Lea 1980; Wang \&
Robertson 1984). This enhances the transport of matter toward the NS
and therefore strengthens the reliability of our assumption.

\end{document}